\documentclass[dvipdfmx]{pasj01}

\Received{$\langle$reception date$\rangle$}
\Accepted{$\langle$acception date$\rangle$}
\Published{$\langle$publication date$\rangle$}

\usepackage{color}
\RequirePackage{ulem}

\usepackage{lineno}

\usepackage[dvipsnames]{xcolor}

\def\aap{A\&A}
\def\apj{ApJ}
\def\apjl{ApJ}
\def\apjs{ApJS}
\def\mnras{MNRAS}
\def\pasj{PASJ}
\def\aj{AJ}

\def\msun{\thinspace M_\odot}
\def\blue#1{\textcolor{blue}{#1}}
\def\blued#1{\textcolor{blue}{#1}}
\def\olived#1{\textcolor{olive}{#1}}

\def\red#1{\textcolor{black}{#1}}

\def\reded#1{\textcolor{red}{#1}}

\def\abra#1#2{[{\rm {#1}} / {\rm{#2}}]}

\def\abranc#1{[{\rm {#1}} / {\rm Fe} / {\rm C}]}
\def\abrancn#1{[{\rm {#1}} / {\rm Fe} / {\rm C (+N)}]}
\def\feoh{[{\rm Fe}/{\rm H}]}

\def\sps{$s$-process}

\def\spsn{$s$-process nucleosynthesis}

\def\mhyph{\  \mathchar`- \, }
\def\dmix{\delta_{\rm mix}}
\def\dtmix{\Delta t_{\rm mix}}
\def\nuc#1#2{${}^{#1}${#2}}
\def\nucm#1#2{{}^{#1}{\rm #2}}
\def\ntauexp{\tau_{\rm ex}}
\def\ntauexpt{\tau_{\rm ex, t}}
\def\ntauexptp{\tau_{\rm ex, t, p}}
\def\ntauexpa{\tau_{\rm ex, ac}}

\def\DeltaCNoH{{\mit \Delta}[{\rm C (+N)}/{\rm H}]}

\def\gneq{\thinmuskip=2mu\smash{\lower2pt\hbox{$\buildrel > \over \sim$}} \thinmuskip=2mu} 
\def\lneq{\thinmuskip=2mu\smash{\lower2pt\hbox{$\buildrel < \over \sim$}}\thinmuskip=2mu}

\def\lesssim{\lneq}
\def\gtrsim{\gneq}

\date{Accepted XXX. Received YYY; in original form ZZZ}

\begin{document}

\title{A comparative study on three modes of s-process nucleosynthesis in extremely metal-poor AGB stars}

\author{Shimako \textsc{Yamada}\altaffilmark{1}, Takuma \textsc{Suda}\altaffilmark{2,3}, Yutaka \textsc{Komiya}\altaffilmark{3}, Masayuki \textsc{Aikawa}\altaffilmark{4} and Masayuki Y. \textsc{Fujimoto}\altaffilmark{1}}%
\altaffiltext{1}{Graduate School of Science, Hokkaido University, Kita 10 Nishi 8, Kita-ku, Sapporo 060-0810, Japan}
\altaffiltext{2}{Department of Liberal Arts, Tokyo University of Technology Nishi Kamata 5-23-22, Ota-ku, Tokyo 144-8535, Japan}
\altaffiltext{3}{Research Center for the Early Universe, University of Tokyo, Hongo 7-3-1, Bunkyo-ku, 113-0033, Tokyo, Japan}
\altaffiltext{4}{Faculty of Science, Hokkaido University, Sapporo 060-0810, Japan}

\email{yamada@astro1.sci.hokudai.ac.jp}


\maketitle

\begin{abstract}
Carbon-enhanced metal-poor (CEMP) stars in the Galactic halo have a wide range of neutron-capture element abundance patterns.
To identify their origin, we investigated three modes of $s$-process nucleosynthesis that have been proposed to operate in extremely metal-poor (EMP) \red{Asymptotic Giant Branch (AGB)} stars: 
    the convective \nuc{13}{C} burning, which occurs when hydrogen is engulfed by the helium flash convection in low-mass AGB stars, 
    the convective \nuc{22}{Ne} burning, which occurs in the helium flash convection of intermediate-mass AGB stars, 
    and the radiative \nuc{13}{C} burning, which occurs in the $^{13}$C pocket that is formed during the inter-pulse periods.
We show that the production of $s$-process elements per iron seed (\sps\ efficiency) does not depend on metallicity below $\abra{Fe}{H} = -2$, because \nuc{16}{O} in the helium zone dominates the neutron poison.  
   The convective \nuc{13}{C} mode can produce a variety of $s$-process efficiencies for  Sr, Ba and Pb, including the maxima observed among CEMP stars. 
The \nuc{22}{Ne} mode only produce the lowest end of \sps\ efficiencies among \red{CEMP models}.
We show that the combination of these two modes can explain the full range of \red{observed enrichment of \sps\ elements} in CEMP stars.
\red{In contrast, the \nuc{13}{C} pocket mode can hardly explain the high level of enrichment observed in some CEMP stars}, even if considering star-to-star variations of the mass of the \nuc{13}{C} pocket. 
   These results provide a basis \red{for} discussing the binary mass transfer origin of CEMP stars and their subgroups.  
\end{abstract}



\section{Introduction}

Extremely metal-poor (EMP) stars are important objects for understanding the chemical evolution and star formation history of the early Universe and the early Galaxy.  
   In the past two decades, the number of known EMP stars has increased significantly thanks to large-scale surveys of Galactic halo stars, such as the HK survey \citep{beers92} and the HES survey \citep{christlieb08}, and follow-up, high-resolution spectroscopic observations with large telescopes.  
A large portion of EMP stars show enhanced carbon abundances relative to iron, with levels ranging up to three orders of magnitude. The fraction of EMP stars that are carbon-enhanced is estimated to be 20-30\% or more for metallicities $\feoh \lesssim -2.5$. (e.g., \cite{lee14} and the references therein).  
    These stars are defined as carbon-enhanced metal-poor (CEMP) stars \citep{beers05}.


CEMP stars are classified into two subtypes: CEMP-s and CEMP-no. CEMP-s stars are enriched in slow neutron-capture process (s-process) elements, while CEMP-no stars are not. The dividing line between the two subtypes is the [Ba/Fe] ratio, with CEMP-s stars having $\abra{Ba}{Fe}\ge 0.5$ and CEMP-no stars having $\abra{Ba}{Fe}<0.5$ \citep{aoki02a,ryan05}.
\red{It is now widely accepted that the \sps\ nucleosynthesis in AGB stars is responsible for the enhancement of carbon and \sps\ in CEMP-s stars}. The s-process elements are produced in the AGB star's outer envelope, and are then transferred to the companion star through stellar winds or binary interactions.
The origin of CEMP-no stars is still debated.
We have hypothesized that CEMP-s and CEMP-no stars originate from binary systems with low-mass and high-mass AGB primary stars, respectively, based on the evolutionary characteristics of EMP stars. \citep{suda04,komiya07}.
It is also proposed that they are formed from the ejecta of peculiar supernovae, such as those produced by the mixing and fallback scenario \citep{umeda02,umeda05,tominaga07,tominaga14}.
It is also believed that they are formed from gas ejected by rapidly rotating massive stars, \red{which have undergone the rotation-induced elemental mixing, leading to the enrichment of carbon and nitrogen} \citep{meynet06, meynet10, choplin16}.

On the other hand, radial velocity monitoring studies have suggested that CEMP-s and CEMP-no stars belong to statistically different populations in terms of binary properties \citep{starkenburg14, hansen16a}.
\red{This implies that CEMP-no stars may belong to longer period binaries compared to CEMP-s stars \citep{suda04,norris13}. } 
   In fact, some CEMP-no stars have been found to have signatures of contamination with s-process material,  (CEMP-low-s stars; \cite{masseron10,spite14b}), which suggests that they may have received mass from a defunct AGB star in a binary system.
  In particular, it has been confirmed that the CEMP-no star HE0107-5240 is in a binary system with a period of 10,000-30,000 days \citep{arentsen19,bonifacio20}, as predicted by  \citet{suda04}.
Furthermore, CEMP-s and CEMP-no stars have been observed to have both common and different features in their abundances and metallicity distributions (e.g., \cite{aoki07,allen12}). Any theory of their origin must be able to explain the scarcity of CEMP-s stars at low metallicities ($\feoh \lesssim -3.5$), while CEMP-no stars continue to exist at lower metallicities. 
  It is hypothesized that these differences between CEMP-s and CEMP-no stars may be related to the formation history of low-mass stars and binaries in the early Universe. In particular, it has been suggested that long-period binaries may be more common at $\feoh \lesssim -3.5$, as \citet{komiya20} predicted.


We are revisiting the origin of CEMP stars by studying the characteristics of the abundances of neutron capture elements. To do this, it is essential to understand the basic properties of the s-process in the early Universe and to develop a solid understanding of \red{the mechanisms that cause the observed variations in the abundances of heavy neutron-capture elements, which are often accompanied by carbon enhancements} \citep{masseron10}.  

There have been many attempts to understand the s-process nucleosynthesis in EMP AGB stars by comparing their observations with theoretical models. 
   \citet{aoki01} were the first to estimate the neutron exposure from the observed enrichment of Sr through Pb in some CEMP-s stars.  
\citet{ma07} compared the s-process materials produced by a single irradiation of neutrons in the helium convective zone with the observed abundances in some CEMP-s stars.  
Bisterzo et al. (2010; 2011) tried to interpret the abundance patterns of individual CEMP-s stars by the materials produced by the s-process nucleosynthesis in the $^{13}$C pocket during the inter-flash in low-mass AGB stars \citep{gallino98}.   
\citet{lugaro12} also discussed the nucleosynthesis with the $^{13}$C pocket and the hydrogen-ingestion in the helium-flash convection in AGB stars with [Fe/H] = -2.3 using a post-processing code, and \citet{abate15a} tryied to apply the results to the CEMP-s stars with known orbital periods.  
  \citet{choplin21} computed the $s$-process nucleosynthesis triggered by the hydrogen ingestion at the early phase of thermal pulses in the AGB phase for  $1 \msun$    model with  $\abra{Fe}{H}=-2.5$ using a nuclear network of 1,091 species.
\red{These authors claimed possible agreement with observations, but in reality some models predict carbon overabundances ten times greater than those observed. This suggests that there is still much that we do not understand about the s-process nucleosynthesis in EMP AGB stars.}


In this paper, we explore the basic characteristics and distinctions of the three modes of s-process nucleosynthesis that are expected to operate under extremely metal-poor conditions.
We compare the abundance patterns of carbon, strontium, barium, and lead with the observations. The abundance patterns of the elements between neutron magic numbers are more sensitive to multiple neutron exposures because these elements are much less abundant than the elements with neutron magic numbers.
 We assume that the observed EMP stars are secondary components of binary systems, as suggested by \citet{komiya07}.
The three modes of \sps\ nucleosynthesis are:
\begin{itemize}
\item the s-process nucleosynthesis that is driven by $^{13}$C, which is produced when hydrogen is engulfed by the helium shell flashes in low-mass EMP-AGB stars.
\item the s-process nucleosynthesis that is driven by $^{22}$Ne, which is produced by $\alpha$-captures in \red{the convective zones driven by the helium shell flashes} in intermediate-mass AGB stars.
\item the radiative $^{13}$C burning in the $^{13}$C pocket, which is formed during the inter-pulse periods.
\end{itemize}

   The evolution and nucleosynthesis of AGB stars are currently subject to uncertainties in various physical aspects, especially in the numerical prescriptions of the convection and the elemental mixing.  
We do not make a detailed comparison of the particular nucleosynthetic materials with the abundance of individual stars. 
   Instead, we try to establish the framework to understand the abundance patterns of neutron-capture elements among CEMP stars by considering the basic characteristics of the materials, obtained from our computations, and their correlation with the carbon enhancements.  
This helps to make clear what is needed for further refinement in the theory of AGB evolution and nucleosynthesis.

  Our paper is organized as follows:
  \S~2 discusses the mechanisms and function of the three modes of s-process nucleosynthesis in EMP-AGB stars, and summarizes the method and approximations used to compute the progress of shell flashes and nucleosynthesis. 
\S~3 presents the results of the computations and discusses the basic properties of s-process nucleosynthesis.   
   \S~4 explores the correlation between the enrichments of the three peak elements, Sr, Ba, and Pb, for each mode.
\S~5 compares our theoretical predictions with the observed abundances in CEMP stars to reveal which mode(s) can reproduce the abundances in CEMP stars.
   \S~6 summarizes the main findings of the paper..

\section{ Method of Computations}

In this section, we first review the three modes of the s-process nucleosynthesis that are proposed to occur in extremely metal-poor AGB stars. We then describe our method for computing the nucleosynthesis during the recurrence of helium shell flashes, and the settings used for each mode. 

\subsection{The characteristics of three modes of \spsn}

\subsubsection{\it The Convective \nuc{13}{C} Burning {\rm (C13C)} Mode}

Hydrogen engulfment can occur during helium shell flashes in low-mass AGB stars with masses of $M \lesssim 3.5 \msun$ and metallicities of $\feoh \lesssim -2.5$ \citep{fujimoto90, fujimoto00, iwamoto04, campbell08, lau09, lugaro09, suda10,choplin22}. 
This is because the entropy barrier in the hydrogen burning shell, which prevents hydrogen engulfment, is lowered by the decreasing mole abundance of CNO elements, or metallicity, in the envelope \red{\citep{fujimoto90}}.
The entropy barrier also depends on stellar mass. Intermediate-mass stars have larger radiation pressure arising from larger gravity than low-mass stars. \red{This raises the entropy barrier, making it more difficult for the helium-flash convective zones to engulf hydrogen from the envelope.}
   
  The hydrogen engulfment occurs in the early stages of the thermal pulse-AGB (TP-AGB) phase. The mixing rate of the engulfed hydrogen increases with the strength of the shell flashes, as the shell flashes grow stronger during the recurrence. 
When the mixing rate becomes sufficiently large, a hydrogen shell flash develops in the middle of the helium convective zone, triggering helium flash-driven deep mixing (He-FDDM) \citep{hollowell90,fujimoto00}
  or called ``dual flashes''  \citep{campbell08}.
This process enriches the envelope with carbon and s-process heavy elements is called the C13C mode. 
   While the mixing rate of hydrogen remains small, the nuclear products are stored in the helium zone and eventually carried out to the surface later by He-FDDM and/or the third dredge-up (TDU) after the helium shell flash grows strong.    
The process of hydrogen engulfment and dredge-up is schematically illustrated in Figure 1 of Nishimura et al. (2009; referred to as Paper~I).

In this study, we treat the mixing rate and amount of hydrogen into the helium zone as parameters. These parameters may vary depending on the strength of shell flashes, the treatment of convection, and the numerical recipe of materials mixing (see the review by \cite{karakas14}).
Specifically, they depend on whether overshooting and/or extra-mixing are considered.
We can constrain these parameters by comparing the resultant materials with the observed abundances. 
   
We specify the amount of mixed hydrogen, $\Delta M_{\rm H, mix}$, in terms of the number ratio, $\dmix$, to \nuc{12}C averaged in the convective zone, defined as; 
\begin{eqnarray}
\dmix = \Delta M_{\rm H, mix} / (M_{\rm He, conv} Y_{12, \rm He, ini}),   
\label{eq:dmix}
\end{eqnarray}
   where \red{$M_{\rm He, conv}$ is the maximum mass of the convective zone during the thermal pulse}, $\Delta M_{\rm H, mix}/M_{\rm He, conv}$ is the mole abundance of mixed hydrogen, and $Y_{12, \rm He, ini}$ is the mole abundance of \nuc{12}{C} in the helium convective zone at the beginning of the hydrogen engulfment. 
In our computations, we mix \nuc{13}{C} instead of hydrogen at the expense of \nuc{12}{C} abundance.
This is because the mixed hydrogen is captured by \nuc{12}{C} in the middle of the convective zone and carried further inwards as \nuc{13}{C} to release neutrons.      
   The duration of mixing, $\dtmix$, is also treated as a parameter, as in Paper~I.  
   The mixing rate is taken to be constant, given by $\dot{Y}_{13,\rm mix} = \dmix Y_{12, \rm He, ini} / \dtmix$. 

\red{
We have explored the nucleosynthesis of neutron-capture processes with maximum neutron densities of up to $\sim 10^{12}$ cm$^{-3}$.
However, this value is not sufficient to follow the hydrogen ingestion by the C13C mode, as other studies have reported neutron densities as high as $> 10^{13}$ cm$^{-3}$ \citep{iwamoto04,cristallo09}.
We will investigate the models with higher neutron densities in a future study.
}  

The He-FDDM increases the CNO abundance in the envelope. The CNO abundance is higher for lower-mass stars because the mass of the helium zone decreases with increasing core mass, while the envelope mass decreases with increasing core mass (e.g., \cite{iwamoto04}).
     The He-FDDM can enrich the envelope with carbon near the observed upper limit of $\abra{C}{H} \simeq 0$, , which can cause the envelope to be blown off by carbon dust driven wind (see \S 3.1).    
   However, for stars with $2\lesssim M/\msun \lesssim 3.5$, the surface carbon enhancement is too low to blow off the envelope after the He-FDDM. This is because the mass in the helium zone is smaller and the mass in the envelope is larger for these stars \citep{suda10}.

Finally, we highlight another potential C13C mode of s-process nucleosynthesis in AGB stars.
This mode may operate during the very late thermal pulse (VLTP), which is the last helium shell flash that occurs while the stars are evolving toward white dwarfs, stripped off the envelope by mass loss \citep{fujimoto77,sugimoto78b,schoenberner79,herwig11}.   
The VLTP can occur regardless of the metallicity and the initial mass because the hydrogen shell burning has already been extinguished.
This rejuvenates the star, creating a red giant with an expanded helium zone.
The VLTP ejects carbon and s-process elements together with the mass of the helium zone by the wind mass loss.
The wind accretion of these products can convert secondary stars to CEMP stars.
Accordingly, the VLTP can be a viable mechanism to form CEMP stars at metallicity of $\feoh >-2.5$.   
However, in order to have the VLTP, the helium shell flash has to be ignited \red{before the star settles into the white dwarf phase}. This limits the frequency of AGB stars that experience VLTP.

\subsubsection{The Convective \nuc{22}Ne\ Burning (C22Ne) Mode}

After the helium flash-driven deep mixing (He-FDDM), the succeeding shell flashes occur without the hydrogen engulfment. This is because the entropy in the hydrogen burning shell grows too large for the helium flash convection to reach the tail of the hydrogen-rich layer.
The possible neutron source for the s-process nucleosynthesis in this case is $\nucm{22}{Ne}(\alpha, n) \nucm{25}{Mg}$.
This mode of s-process nucleosynthesis is called the C22Ne mode, which occurs after the C13C mode or the R13C mode (see the next subsection).
The C22Ne mode enriches the surface carbon abundance through the third dredge-up (TDU).
However, \red{due to the lower efficiency of the production of s-process elements than the C13C mode, the abundance ratio of s-process to carbon materials decreases by the TDU} until the surface carbon abundance increases up to $\abra{C}{H} \simeq 0$.
At this point, the carbon-dust driven wind blows off the envelope (see \S 3.1 for details).


In the helium flash convective zone, \nuc{22}{Ne} is formed by $\alpha$-captures of \nuc{14}{N} in the ash of hydrogen shell burning. 
   The burning of \nuc{22}{Ne} requires a high temperature of $T \gtrsim 3 \times 10^8 \hbox{ K}$, which is only realized during the shell flashes of stars with a large core mass or high mass (e.g., \cite{iben75b,truran77}).    
   The maximum temperature, $T_{\rm He}^{\rm max}$, reached in the helium flash convective zone, can be read as a function of the core mass, and hence, the stellar mass (see Fig~\ref{fig:A2maxtemp} in Appendix).

In this study, we treat the initial abundance of \nuc{14}{N} newly added to the helium convective zone as a parameter. We do this because the TDU increase the CNO abundances in the envelope during the TP-AGB phase, and the carbon abundance is eventually converted to nitrogen abundance because of the CN reaction.  
Therefore we set  the upper limit of the initial abundance of \nuc{14}{N}, $Y_{\rm 14N,ini}$, to be $\simeq Y_{\rm C, \odot}$ which corresponds to the observed upper limit of carbon abundance in CEMP stars.

In this mode, we need to consider the multiple irradiation of neutrons during the recurrent shell flashes, as argued by \citet{ulrich73} and \citet{iben75b}.   
The mass of the helium zone decreases with the stellar mass (e.g., Sugimoto \& Nomoto 1975), so massive AGB stars that can burn \nuc{22}{Ne} must experience many cycles of shell flashes.  
During the recurrence, some of the matter that has been irradiated by neutrons in the convective zone is left over by the TDU and incorporated into the convection of the succeeding shell flash, where it is exposed to neutrons again.  
   \citet{iben75b} argues that this multiple irradiation leads to the abundance patterns of the solar system s-process elements (see also \cite{truran77}).  
However, it is also important to note that neutron poisons, such as oxygen, neon, and magnesium isotopes, are also accumulated during the recurrence and impede the progress of the s-process nucleosynthesis.
In our computations, we will follow the recurrence of shell flashes \red{until the abundance patterns converge due to multiple irradiations.}
We will use the fraction, $r$, of the overlap between the two successive helium flash convective zones as a parameter (hereafter $r$ is the overlapping factor).

\subsubsection{Radiative \nuc{13}{C} Pocket Burning (R13C) Mode}

This mode assumes that hydrogen is injected \citep{gallino98} or partially diffused \citep{goriely00} from the bottom of the hydrogen-rich layer down into the \nuc{12}{C}-rich helium zone, left over by the TDU, to form a thin \nuc{13}{C}-rich layer.   
   In this \nuc{13}{C} pocket or partial mixing zone, \nuc{13}{C} burns to promote the \sps\ nucleosynthesis even before the helium shell flash is ignited  
 \citep{straniero95,straniero97}.   
The timescale of neutron release is regulated by the compressional heating of the helium zone during the inter-flash period, which is much longer than the C13C and C22Ne modes.
The succeeding shell flash involves the zone left over from the preceding shell flash, including the \nuc{13}{C} pocket and the ash of hydrogen shell burning, newly added to the core.  
Then, processed by the nuclear burning in the flash convection, the materials from the pocket are dredged up into the envelope along with the s-process materials newly synthesized via the C22Ne mode.
The formation of the \nuc{13}{C} pocket may repeat as the shell flash recurs, so that we should consider the effect of multiple neutron irradiations and also the accumulation of neutron poisons during the recurrence, as the C22Ne mode did.  

Since the formation mechanisms of the \nuc{13}{C} pocket are not yet fully understood 
(e.g., \cite{karakas14}), a parameterized approach has been necessary to describe both the mass of the \nuc{13}{C} pocket (pocket size) and the amount of mixed \nuc{13}{C}.  
   The pocket size can be specified by the mass ratio of the \nuc{13}{C} pocket to the helium convective zone, i.e., $\zeta_{\rm p} = \Delta M_{\rm 13C, p}/ \Delta M_{\rm He, conv}$.
   The amount of mixed hydrogen is determined by the mixing parameter, $\dmix$, which is given by \red{$\Delta M_{\rm H, mix} / (M_{\rm 13C, p} Y_{12, \rm He, ini})$}. 
   \citet{gallino98} have discussed the parameter ranges that can reproduce the solar system \sps\ abundances (see also \cite{goriely00}), and argue for $\zeta_{\rm p} \approx 1/20$ and $\dmix = 0.03$ (see also Bisterzo et al. 2010).  
   We use these values as typical to study the characteristics of the R13C mode, also exploring the parameter dependences by changing the pocket size and the degree of hydrogen mixing.   


\subsection{Progress of Shell Flashes and Nucleosynthesis}

Our method of computation is the same as that used to compute the nucleosynthesis during the hydrogen shell flashes, induced by the hydrogen injection into the helium core, in RGB stars \citep{fujimoto00, aikawa01} and the helium shell flashes with hydrogen engulfment in EMP AGB stars (Paper~I).
The thermal structure of the helium zone during the shell flashes is solved analytically \citep{sugimoto78}.  
y using this solution, a semi-analytical expression for the evolution of shell flashes in finite amplitude is formulated \citep{fujimoto82a,fujimoto82b}.
This allows us to simultaneously compute the thermal history of the helium flash and the progress of nucleosynthesis in the helium flash convective zone under one-zone approximation with the aid of thin shell approximation \citep{hayashi62}.  
   
\begin{figure}

\includegraphics[width=\columnwidth,bb=0 0 676 720, clip]{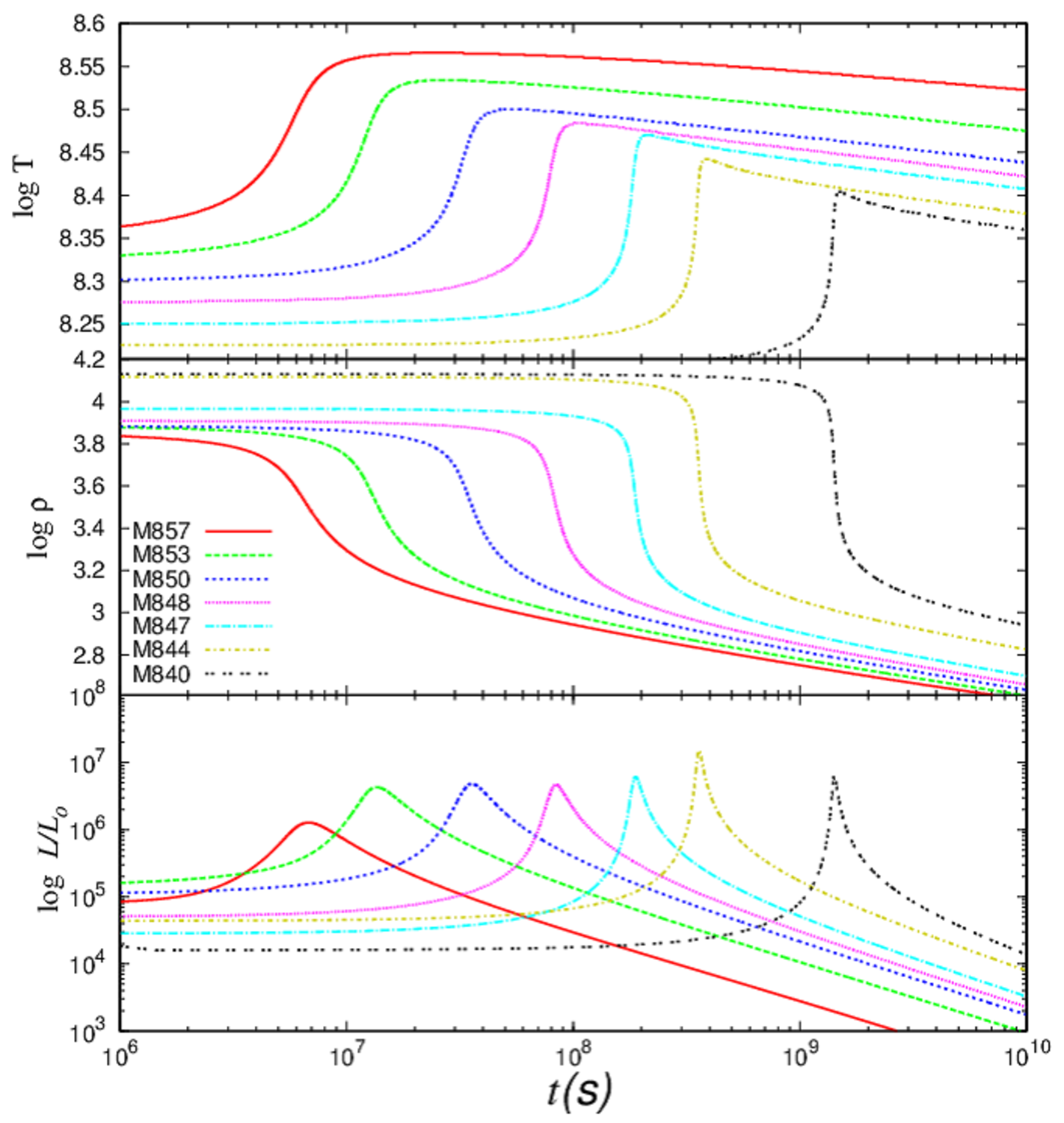}

\caption{
Progress of helium shell flashes for various models; time variations of the temperature (top panel) and density (middle panel) in the bottom of the helium burning shell and the helium burning rate (bottom panel).  
   The model names are designated by the logarithm of the maximum temperature reached during the shell flash, which is shown in the key of the middle panel.  
}
\label{fig:flash}
\end{figure}

We have described the semi-analytical model and the thin shell approximation in Appendix.  
   The strength of helium shell flashes is determined by three parameters: the mass and radius of the core interior to the helium burning shell ($M_c$ and $r_c$), and the proper pressure in the helium burning shell ($P_*$), defined in eq.~(\ref{eq:pHe}). Alternatively, the mass of the helium zone, $M_{\rm He}$, can be used to define the strength of the shell flashes.  
   The model parameters and the characteristics of the resultant helium shell flashes in this work are summarized in Table~\ref{tab:param}.  
   The variations of the temperature and density in the bottom of the helium burning zone and the helium burning rates (which is equivalent to the luminosity emitted by the helium burning) are shown in Figure~\ref{fig:flash}.  
   The model number represents the logarithm of the maximum temperature reached during the shell flashes.
For instance, model 857 means that the maximum temperature is $10^{8.57}$ K).  
   The maximum temperature and density are chosen so as to embody the dependences of the shell flashes on the total mass and metallicity of stars and the thermal state of the core (see Figures~\ref{fig:A1massradius}-\ref{fig:A3coremass} in Appendix). 
   The use of the semi-analytical model allows us to survey a wide range of parameter space systematically.




The nucleosynthesis during the helium shell flashes is computed simultaneously with the thermal history of the helium flash convective zone.
We use the nuclear reaction network from Paper I, which has been extended to include 318 isotopes of 84 element species from \nuc{1}{H} and neutron up to \nuc{210}{Po} with the lifetime longer than 10 days.   
   The nuclear reactions include hydrogen and helium burnings, the proton-, neutron-, and alpha-captures, as well as $\beta$-decays.   
   The reaction rates are taken from \citet{angulo99} and \citet{caughlan88} for charged-particle reactions and from \citet{bao00} for neutron-capture reactions.  
   The reaction rates for light elements of the mass number $A \le 35$ are the same as in \citet{nishimura09}. 
   The cross section of $\nucm{17}{O} (n, \alpha) \nucm{14}{C}$ is set at 10 mb (\cite{koehler91}; cf. Wagemans et al. 2002) in the temperature range appropriate for our problems, and \red{the neutron-capture cross section of $\nucm{17}{O}$ is set equal to that of \nuc{16}{O}}.

\section{Results of the Three modes of \lowercase{$s$}-process nucleosyntheses}

   In this section, we present the results of our computations and discuss the characteristics of the three modes of s-process nucleosynthesis, C13C, C22Ne, and R13C, which were explained earlier, in EMP AGB stars.   

\subsection{The Convective \nuc{13}{C} Burning (C13C) Mode} \label{sec:c13c}

In our computations, we only consider the case where the neutron irradiation occurs once.  
However, the helium convection may engulf the hydrogen in the envelope repeatedly, which could lead to the envelope being enriched with carbon beyond $\abra{C}{H} \simeq -2.5$.  
   In addition, the shell flashes without hydrogen engulfment may follow the He-FDDM until the star loses its envelope due to the stellar wind.
The increase in the carbon abundance via the TDU will reinforce this process. 
The observed upper limit of the carbon abundance, $\abra{C}{H}\simeq 0$, in EMP stars can be explained by the stellar wind driven by carbonaceous dust, which strips the envelope of AGB stars and evolves into white dwarfs (e.g., Woitke 2006; Lagadec \& Zijlstra 2008).

\begin{figure}
\includegraphics[width=\columnwidth,bb=0 0 556 806, clip]{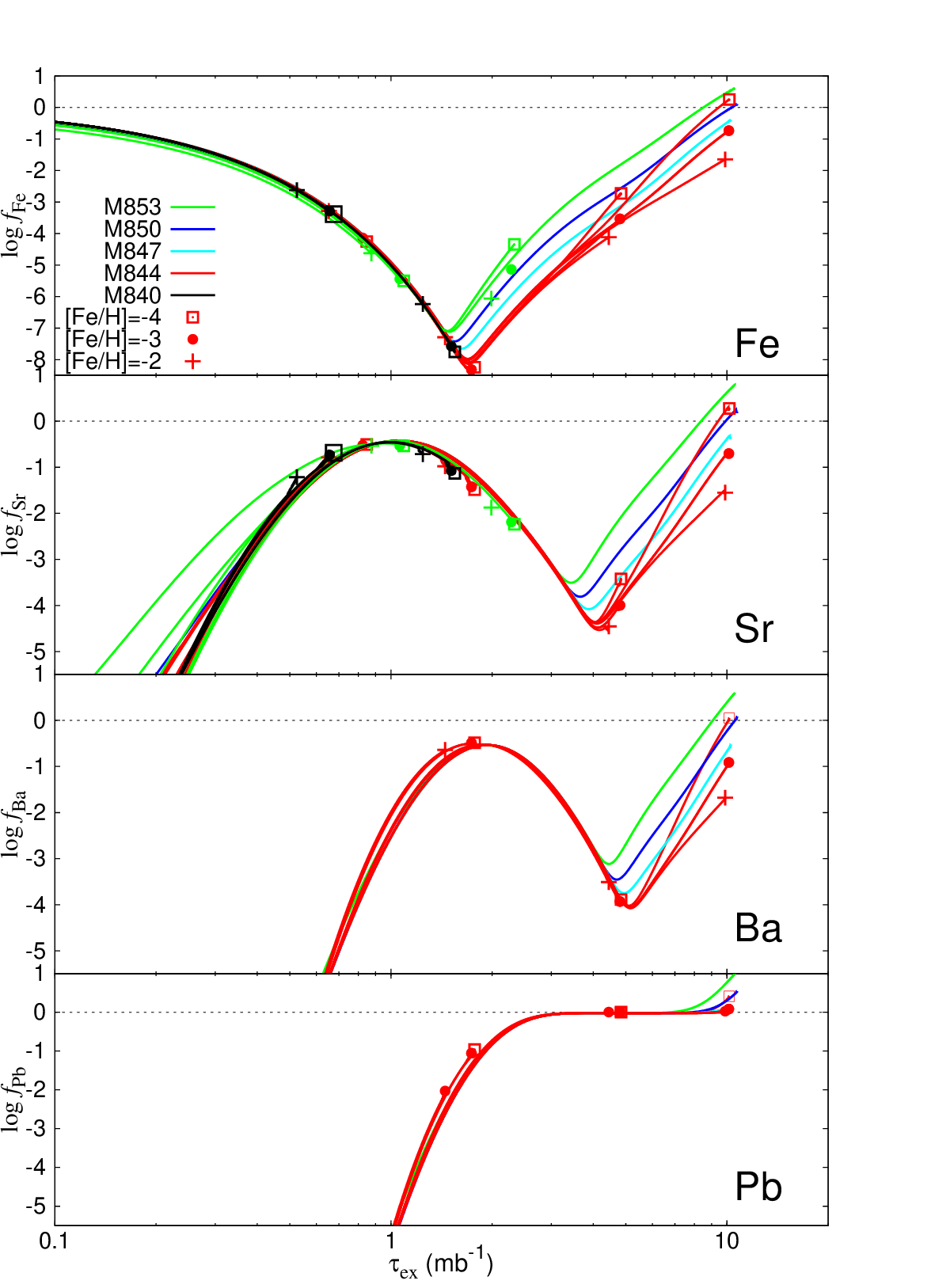}

\caption{ 
Progress of the \sps\ nucleosynthesis of the C13C mode as a function of the neutron exposure, $\ntauexp$, for different amounts of engulfed hydrogen and various metallicities.  
   Plotted are the efficiency factors of Fe, Sr, Ba, and Pb, defined by eq.~(2)  
   The amount of mixed hydrogen is taken to be $\dmix = 0.001$, 0.02, and 0.1 and the duration is set at $\dtmix = 10^8$ sec.  
   Lines (colored) denote the different models of helium shell flashes.  
   Symbols indicate the end points for the different metallicities: $\feoh = -2$ (crosses), $\feoh = -3$ (solid circles) and $-4$ (open squares).
   The three same symbols for each metallicity are arranged from left to right in order of increasing $\dmix$.   }  

\label{fig:tauC13C}
\end{figure}

First, we follow the nucleosynthesis during a shell flash without hydrogen engulfment to derive the compositions in the helium convective zone.
Then, we compute the succeeding shell flash with hydrogen engulfment taken into account.  
We define the efficiency factor, $f_{\rm Z}$, of \sps\ nucleosynthesis as the ratio of the mole abundances of element Z, $\rm Y_{\rm Z}$, to that of the pristine or initial iron, $\rm Y_{\rm Fe,p}$.
\begin{eqnarray}
f_{\rm Z}=\rm Y_{\rm Z}/Y_{\rm Fe,p}
\label{eq:efffactor}
\end{eqnarray}
In other words, the efficiency of \sps\ nucleosynthesis is measured by the conversion rate of the pristine iron to element Z.
In Figure~\ref{fig:tauC13C}, the efficiency factor, $f_Z$, is plotted against the neutron exposure, $\tau_{\rm ex}$.
  The values of $f_Z$ are represented by a nearly single track as a function of \blue{$\tau_{\rm ex}$}, regardless of the models and the mixing parameters. 
However, there is a diversity at low $\ntauexp (\ll 1)$, for Sr abundance only in the models of high temperatures.
This is due to the contribution of the $\nucm{22}{Ne} (\alpha, n) \nucm{25}{Mg}$ reaction.
   A small offset also arises from the competition with the intervening $\beta$-unstable nuclides.  
The first \sps\ peak element, Sr, reaches its local maximum of $f_{\rm Sr, max} = 0.32$ at $\ntauexp = 1.08 \ {\rm mb}^{-1} (\dmix \simeq 0.001)$, when the second peak element, Ba, still remains very small as $f_{\rm Ba} = 0.006$.  
The Ba abundance attains its local maximum, $f_{\rm Ba, max} = 0.26$, at $\ntauexp = 1.95 \ {\rm mb}^{-1} (\dmix \simeq 0.003)$, when Sr decreases to $f_{\rm Sr} = 0.05$.  
   For the exposure of $\ntauexp \gtrsim 3\ {\rm mb}^{-1}$, almost all the pristine irons are converted to the third peak element, Pb.

The first phase of \sps\ nucleosynthesis is the iron seed phase, where most neutrons are absorbed by iron seed nuclei. After the consumption of almost all iron-group elements, the neutron recycling reactions between O and Ne/Mg become the dominant neutron-capture process. This is the Ne/Mg seed phase. The reason for the increase of Fe, Sr, and Ba again is the production of iron seed nuclei again by the neutron-capture reactions from Ne/Mg seeds.
We define the Fe seed phase as the period from the beginning of the nucleosynthesis until the amount of heavy Z element synthesized through neutron capture of Fe seed is exceeded by that of Ne and Mg seeds. The Ne/Mg seed phase begins after this point. The transition from the Fe seed phase to the Ne/Mg seed phase occurs at the upturn after the $f_{\rm Z}$ hitting the bottom, as seen in the top panel of Figure 2.
At the transition to the Ne/Mg-seed phase, $f_{\rm Sr}$ and $f_{\rm Ba}$ hit the bottom, 4.0 and 3.5 dex below their maxima on the Fe-seed phase at $\ntauexp \simeq 4.2$ and $5.2 \hbox{ mb}^{-1}$ in Model M840, respectively.  
For the shell flashes with higher temperatures, the upturns occur at smaller neutron exposures with larger efficiency factors because of larger production of Mg isotopes by $\alpha$-capture reactions.

At first, the amount of Ne/Mg seeds is proportional to the pristine metallicity. However, the increase in the CNO elements in the envelope through HeFDDM or TDUs allows the Ne/Mg seeds to be formed by \nuc{14}{N} descendants of the CNO elements in the envelope.
   After the hydrogen mixing starts, the isotopes of the Ne/Mg seeds form from oxygen via neutron and $\alpha$-captures. The amount of the Ne/Mg seeds is then determined by the mixing degree of hydrogen, irrespective of the metallicity. In other words, the amount of heavy element which is synthesized by neutron capture of Ne/Mg seeds is the same, irrespective of the metallicity, if the temperature and the amount of mixed hydrogen are the same. Therefore, the efficiency factors tend to be inversely proportional to the pristine metallicity. This can be seen in the red solid lines in the panels for Fe, Sr, and Ba in Fig.~\ref{fig:tauC13C}

\begin{figure}
\includegraphics[width=\columnwidth,bb=0 0 450 300, clip]{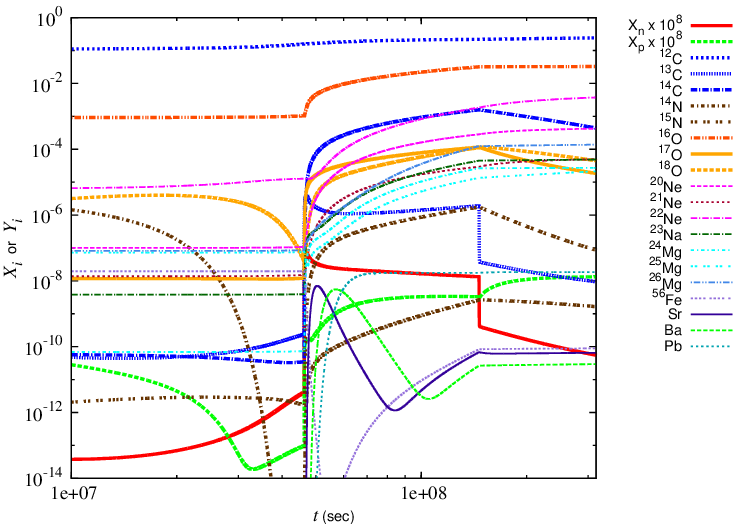}
\includegraphics[width=\columnwidth,bb=0 0 450 300, clip]{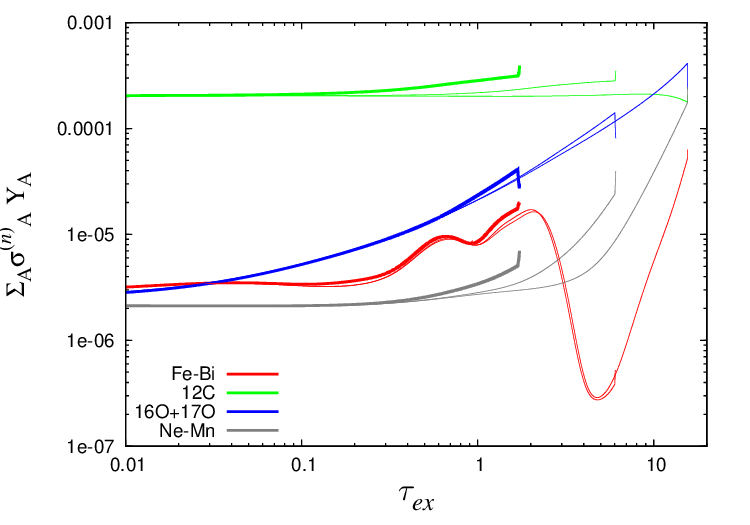}

\caption{
   {\sl Upper panel}: The progress of nucleosynthesis, i.e., the time variation of the abundances, $X_{A}$, for light elements, $X_{n}$ and $X_p$ for neutron and proton respectively and the mole abundance, $Y_i$, for iron and heavier elements (the sum of all stable isotopes and all parent unstable isobars)
in the helium convective zone for a shell flash of Model M844 with the hydrogen engulfment of $\dmix = 0.03$ and the metallicity $\feoh = -3$;  
   the hydrogen engulfment starts at the peak of helium burning rate and lasts for an interval of $\dtmix = 10^8$ sec at constant rate. {\sl Lower panel}: 
   The neutron absorption coefficients, $\Sigma_{\rm A} \sigma_{\rm A}{}^{(n)}\rm{Y}_{\rm A}$, i.e., the products of the neutron absorption cross sections, $\sigma_{(n)} {}^{\rm A}$, 
 and the mole abundance, $Y_A$, of nuclides with the mass number $A$, are compared for \nuc{12}{C}, the sums of \nuc{16}{O} and \nuc{17}{O}, the light elements ($20 \le A \le 55$) and the heavy elements ($A\ge 56$) in the helium convective zone for the same model
   as in upper panel but for the metallicity $\feoh = -2$.
  \red{Three lines with the same color represent models with different  $\dmix = 0.003$, $0.03$, and $0.3$, where the end points of s-process nucleosynthesis lie from left to right in order of increasing $\dmix$.}

}
\label{fig:ysigma-tau_cc13-2} 
\label{fig:SYC13C-2}
\end{figure}


The \sps\ nucleosynthesis in extremely metal-poor (EMP) stars is independent of metallicity. This is because oxygen is the second most abundant metal after carbon, and it is also the most efficient neutron poison. (Paper~I, see also \cite{gallino10}).

  Figure~\ref{fig:ysigma-tau_cc13-2} shows the progress of the abundances and the neutron absorption rate for the major elements in the helium convective zone. As can be seen from the upper panel, oxygen is the second most abundant metal after carbon, and it increases further due to the neutron-producing reaction $\nucm{13}{C} (\alpha, n) \nucm{16}{O}$.   
As can be seen from the lower panel, oxygen dominates other elements as the neutron poison. This is because of the neutron recycling chain of $\nucm{12}{C} (n,\gamma) \nucm{13}{C} (\alpha, n) \nucm{16}{O}$ \citep{gallino88,travaglio96} and the $\nucm{12}{C} (\alpha, \gamma) \nucm{16}{O}$ reaction, which produce more oxygen than mixed \nuc{13}{C} (see Paper~I).

Initially, for $\feoh = -2$, the neutron absorption rate competes between oxygen and heavy elements (iron and its progeny). However, as the \sps\ nucleosynthesis progresses, oxygen becomes the dominant absorber of neutrons. 
   Neutron absorption by \nuc{16}{O} is rapidly followed by the $\nucm{17}{O} (n, \alpha) \nucm{14}{C}$ reaction due to its larger cross section. The isotopic ratio of oxygen reaches a steady state, where $Y_{17O}/Y_{16O} \simeq \sigma_{n\gamma,16} / \sigma_{n\alpha,17} (= 0.0038)$.
 Therefore the absorption rate of \nuc{16}{O} is equivalent to that of \nuc{17}{O}.

The neutron absorption rate of heavy elements (indicated by red lines in the lower panel of Fig.~\ref{fig:ysigma-tau_cc13-2}) also increases with neutron exposure but drops sharply when abundant \nuc{208}{Pb} begins to act as bottleneck for neutron absorption.
   However, it increases again as heavy elements are synthesized through neutron-capture reactions starting from Ne/Mg isotopes.
It is to be noted that the neutron absorption rate by heavy elements is always smaller than that of oxygen and light elements from Ne to Mn.
Additionally, the neutron absorption rate by Ne and Mg, remains lower than that of oxygen throughout the shell flash in models with low temperatures.
As a result, in the metal-poor conditions of $\feoh \lesssim -2$, the maximum neutron density and neutron exposure are primarily influenced by the neutron poison, such as oxygen, as neutron absorption hinders the increase in neutron exposure.

\begin{figure}
\includegraphics[width=\columnwidth,bb=0 0 410 300, clip]{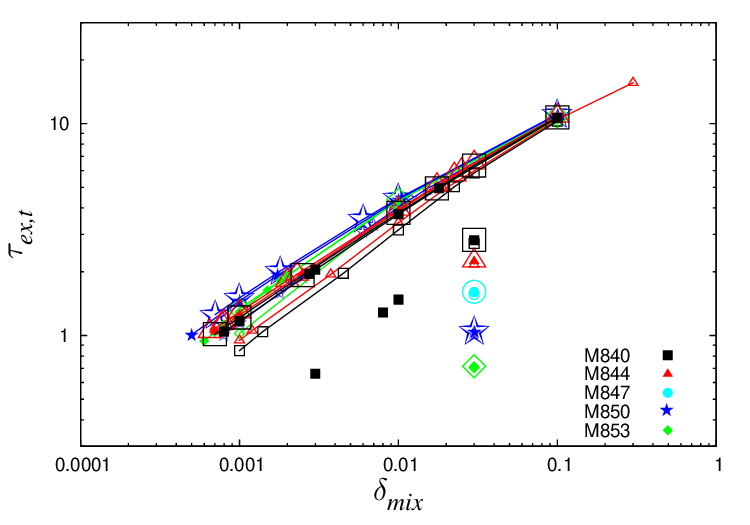}

\caption{ 
The total neutron exposure, $\ntauexpt$, as a function of the mixing parameter, $\dmix$, for the C13C mode (symbols with lines) and for the \nuc{13}{C} burning in the pocket (separate symbols).  
   Symbols denote the models, as given in the key, and the metallicity, $\feoh = -4$ (large open), $-3$ (filled), and $-2$ (small open), respectively. 
}
\label{fig:dmix-tau} 
\end{figure}

Figure~\ref{fig:dmix-tau} shows the total neutron exposure, $\ntauexpt$, reached in the convective zone as a function of the mixing parameter, $\dmix$, for the various models with different metallicities.
The resultant $\ntauexpt$ values are almost the same for the models with $\feoh = -3$ and $-4$.
However, for $\feoh = -2$, the $\ntauexpt$ values are slightly smaller due to the contribution of heavy elements.   
   This difference is a few percent for the models with low temperatures and small $\dmix$, but it decreases for the models with high temperatures and large $\dmix$, as the abundances of oxygen and \red{light elements with $20 \leq A \leq 55$} increase.  

The neutron density, $n_n$, is the result of the balance between the mixing rate, $\dot{Y}_{13, \rm mix}$,  and the destruction rate of neutrons, which is the sum of the neutron absorption rates apart from the recycling reactions.
   This means that the total neutron exposure reached during the mixing can be estimated as follows: 
\begin{eqnarray} 
   \ntauexpt \approx \frac{ Y_{\rm 12, He, ini} \ \dmix}{\left \langle \blue{2 \cdot}  \sigma_{n\gamma,16} Y_{16}   + \sum_{\rm Ne, Mg} \sigma_{n\gamma, A} Y_A \right \rangle},
\label{eq:ndss} \label{eq:ntauexpt} 
\end{eqnarray}
The denominator of this equation gives an average of the neutron absorption rates, with the small contribution from the heavy elements neglected.  
The effect of the absorption by $\nucm{17}{O}$ is simply to double the absorption rate of $\nucm{16}{O}$.
Since the neutron absorption by oxygen and light elements is proportional to the neutron exposure for $\sigma_{n\gamma,16} \ntauexp \ll 1$ (see Appendix 1 of Paper~I), from eq.~(\ref{eq:ntauexpt}), $\ntauexpt$ can be expressed as $\ntauexpt \propto \dmix^{1/2}$, which is also seen in Fig.~\ref{fig:dmix-tau}.
  
For higher temperatures, neutrons are released additionally by the \nuc{22}{Ne} burning.
However, Mg isotopes are also produced simultaneously via the $\alpha$-capture of Ne isotopes, which act as effective neutron poisons.  
Therefore, $\ntauexpt$ reaches a ceiling for Model M850 and decreases for the models of even higher temperatures.
On the other hand, the $\nucm{17}{O}(\alpha, n)\nucm{20}{Ne}$ reaction does not act as a neutron source for high temperatures, since the reaction rate of $\nucm{17}{O}(n, \alpha)\nucm{14}{C}$ is much larger than that of $\nucm{17}{O}(\alpha, n)\nucm{20}{Ne}$ at $T=3 \times 10^8$K.
   Finally, $\ntauexpt$ grows independently of temperature and metallicity for sufficiently large $\dmix$, as the neutron captures overwhelm the $\alpha$-captures among O and Ne/Mg isotopes.    

In summary, the C13C mode can be divided into two phases, depending on the mixing parameter or the neutron exposure. 
The first phase is the Fe seed phase. In this phase, the pristine iron-group elements act as seeds for the production of \sps\ elements.
The efficiency factors for Sr and Ba increase and approach a peak, and then start to decrease as the pristine iron-group elements are converted to Pb.
The second phase is the Ne/Mg seed phase. This phase begins when the efficiency factor hits a minimum and then starts to increase again ($\tau_{\rm ex} > 3.5 \sim 5 \hbox{ mb}^{-1}$).
In this phase, neon and magnesium isotopes act as seeds for the production of s-process elements, triggered by the $\alpha$-captures of \nuc{14}{N} and \nuc{16}{O} in the helium zone.
In EMP stars with $\feoh \lesssim -2$, the progress of the s-process nucleosynthesis is relatively insensitive to the pristine metallicity. Therefore, the efficiency factors are independent of the metallicity in the Fe seed phase, but inversely proportional to the metallicity in the Ne/Mg seed phase.
This can be seen in Fig.~\ref{fig:tauC13C}.

\subsection{The Convective \nuc{22}{Ne} Burning (C22Ne) Mode}\label{sec:c22ne}

In this mode, the number of neutrons released in the helium convective zone increases as the abundance of \nuc{14}{N} increases in the helium convective zone due to the third dredge-up (TDU) during the TP-AGB phase.
\red{The initial abundance ratio of \nuc{14}{N} to \nuc{12}{C} in the helium convective zone, which is referred to as $\delta_{\rm 14N}$, can be written as: }
\begin{eqnarray}
   \delta_{\rm 14N} & \equiv & Y_{\rm 14N, ini} / Y_{\rm 12, He, ini} \nonumber \\
   & = & 0.01 (Y_{\rm CNO, env}/Y_{\rm C, \odot})  (0.3/X_{\rm 12,He,ini}),  
\label{eq:N_He-abundance}
\end{eqnarray} 
   where $Y_{\rm CNO, env}$ is the mole abundance of CNO elements in the envelope and  $X_{\rm 12,He,ini}$ is the helium abundance in the helium zone (typically 0.3). 
This ratio can be compared to the mixing parameter, $\dmix$, in the C13C mode as an indicator of the number of neutrons released.
However, not all of the \nuc{22}{Ne} will burn and release neutrons.
Additionally, this mode produces the strong neutron poison \nuc{25}{Mg}.
Therefore, the efficiency of s-process nucleosynthesis in this mode is greatly depressed compared to the C13C mode, as stated by \citet{iben75b}.  

\begin{figure}
\includegraphics[width=\columnwidth,bb=0 0 380 300, clip]{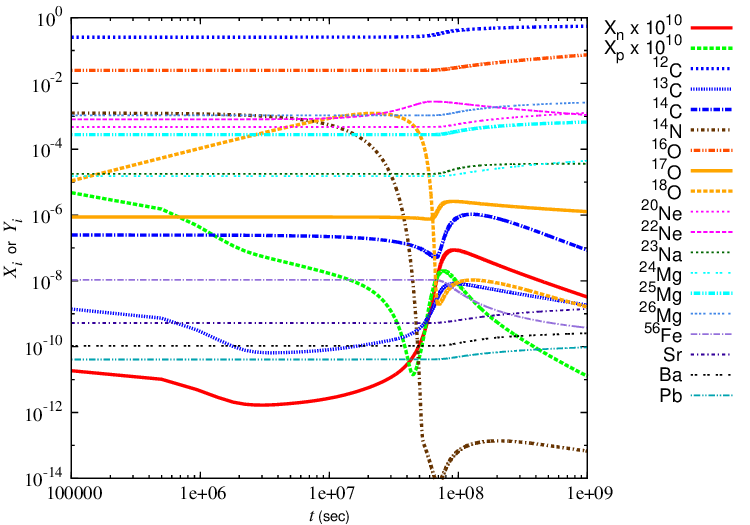}
\includegraphics[width=\columnwidth,bb=0 0 430 300, clip]{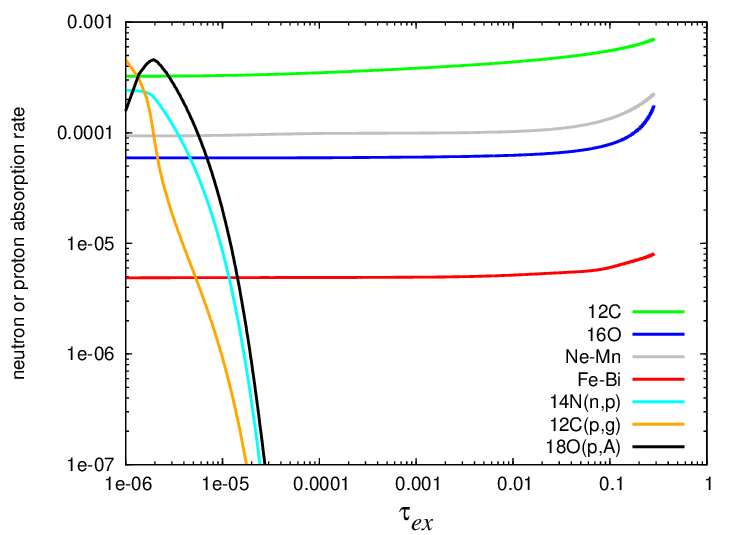}

\caption{ 
Progress of nucleosynthesis against time (upper panel) and variation of the neutron and proton absorption rates, $\Sigma_{\rm A} \sigma_{\rm A}{}^{(n)}\rm{Y}_{\rm A}$ or  $\Sigma_{\rm A} \sigma_{\rm A}{}^{(p)}\rm{Y}_{\rm A}  (\rm{Y}_{\rm p}/ \rm{Y}_{\rm n})$, against neutron exposure (lower panel) for the C22Ne mode during the 12th shell flash of Model M850 with the initial abundance of \nuc{14}{N}, $Y_{\rm 14N, ini} = Y_{\rm C,}{}_\odot$, and the overlapping factor, $r = 0.5$ for the metallicity, $\feoh = -2$. 
   The same as in Fig.~\ref{fig:ysigma-tau_cc13-2}, but $\sigma_{\rm A}{}^{(p)}$ denotes the proton absorption cross sections via $(p, \gamma)$ \blued{or} $(p, \alpha)$ reactions.   
}
\label{fig:22Ne-progress}
\end{figure}
%
We calculate the materials that are repeatedly irradiated by neutrons during the recurrent shell flashes. 
  The yield abundance \red{converges to certain abundance pattern} around the 12th flash for Model M850 with $Y_{\rm N, ini} = Y_{\rm C, \odot}$.
\red{This asymptotic s-process abundance distribution} has been shown by Truran \& Iben (1977; see their Fig.~5), although it is achieved slightly later in our calculations than theirs because of the lack of heavy elements that act as neutron poisons during the recurrence.
Figure~\ref{fig:22Ne-progress} shows the progress of nucleosynthesis and the neutron absorption for the major constituents during the 12th shell flash for Model M850 with $Y_{\rm N, ini} = Y_{\rm C, \odot}$.   
   As seen from the upper panel, the burning of \nuc{22}{Ne} becomes active as the temperature rises to the peak at  $t \simeq 4 \times 10^7$ (s) (see Fig.~\ref{fig:flash}). 
Neutrons liberated by \nuc{22}{Ne} burning are mostly absorbed by \nuc{12}{C} to form \nuc{13}{C}, which enters into the recycling chain, as stated above for the C13C mode, until they are absorbed by \nuc{16}{O} and \nuc{25}{Mg}.  

The neutron exposure during the flash is evaluated from the balance between the production rate and destruction rate of neutrons. The total neutron exposure is evaluated by the following equation:
\begin{eqnarray}
   \ntauexpt \simeq \int \frac {\xi_{\rm 22Ne} / (1 + \xi _{\rm 22Ne})} { \olived{2 \cdot}\sigma_{n\gamma,16} Y_{16}+ \sum_{\rm Ne, Mg} \sigma_{n\gamma, A} Y_A  } d Y_{22}, 
\label{eq:tau_N22}
\end{eqnarray}
   where $\xi_{\rm 22Ne}$ denotes the branching ratio of $(\alpha, n) $ to $ (\alpha, \gamma) $ reactions.  
During the recurrent flashes, Ne and Mg isotopes, especially \nuc{25}{Mg}, are accumulated to be much more abundant than in the C13C mode.
These isotopes act as neutron poisons, which prevent the neutron exposure from becoming large. 
Therefore, the neutron exposure remains small, e.g., $\ntauexpt = 0.22 \hbox{ mb}^{-1}$, while 57 \% of \nuc{22}{Ne} burns during the shell flash in this model.   
The neutron exposure increases only logarithmically for greater burning of \nuc{22}{Ne} and larger $Y_{\rm 14N, ini}$, as the poisons increase in proportion.   

\red{As seen in the lower panel of Fig.~\ref{fig:22Ne-progress}}, the neutron absorption rate of \red{the light elements from Ne to Mn} is comparable to or more than that of oxygen.
At high metallicities of $\feoh \gtrsim - 1$, the pristine iron and its progeny catch up and overwhelm these light elements in neutron absorption.
Also plotted is the neutron absorption rate by the $\nucm{14}{N} (n, p) \nucm{14}{C}$ reaction.  
This reaction is much faster than the neutron capture of \nuc{16}{O}, and it makes the proton density much larger than in the early stage.  
If liberated protons are captured by \nuc{12}{C}, they revert to neutrons again via the reactions $\nucm{12}{C} (p, \gamma) \nucm{13}{N} (e^+ \nu) \nucm{13}{C} (\alpha, n) \nucm{16}{O}$. 
However, if they are captured by \nuc{18}{O}, for which the $(p, \alpha)$ rate is greater by a factor of $\sim 10^3$ than the $(p, \gamma)$ rate of \nuc{12}{C} \citep{angulo99}, protons will be consumed and neutrons will be finally removed.   
Therefore, \nuc{14}{N} is effective as a neutron poison only when the proton capture rate of \nuc{18}{O} is comparable to or larger than the neutron capture rates by \red{oxygen and light elements from Ne to Mn}.  
   In the early stage, \nuc{14}{N} is the dominant neutron poison. 
However, its contribution to the neutron poison is small throughout the shell flash because \nuc{14}{N} is depleted by $\alpha$-capture well before the temperature rises high enough to burn \nuc{22}{Ne}.  
The same is true for the C13C mode, since hydrogen engulfment does not occur until the helium burning rate grows large, at which point \nuc{14}{N} is largely depleted (see Fig~\ref{fig:SYC13C-2} (top panel)).

\begin{figure}
\includegraphics[width=0.8\columnwidth,bb=-50 0 686 1095,clip]{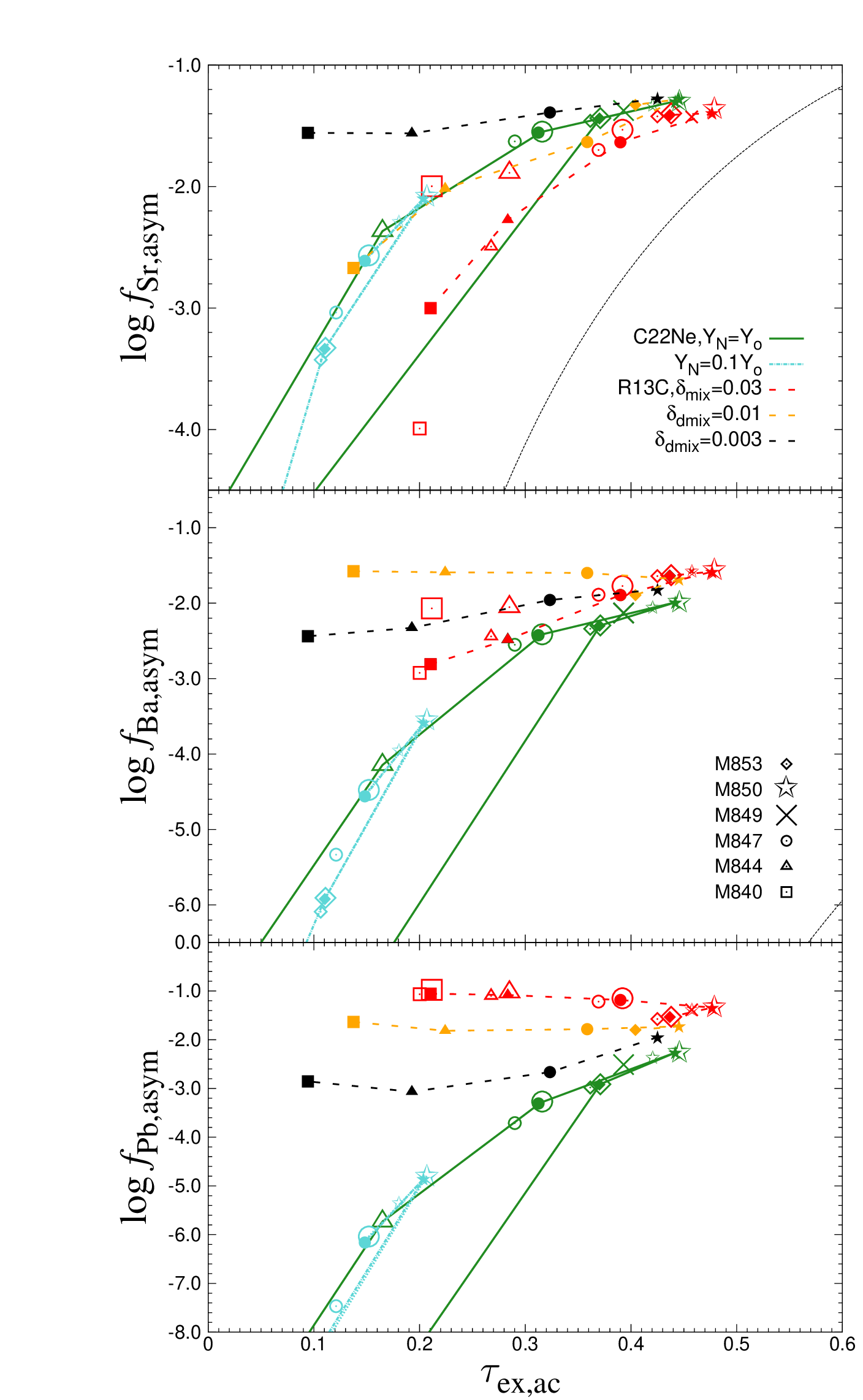}

\caption{
\red{The asymptotic efficiency factors, $f_{\rm Sr, Ba, Pb,asymp}$, of the C22Ne and R13C modes for Sr, Ba, and Pb as a function of  the accumulative neutron exposure. The accumulative neutron exposure, $\ntauexpa$, is defined by eq.~(\ref{eq:nexpac}). }   
\red{ Symbols for the maximum temperatures in the helium flash convection and the metallicities are the same as in Fig.~\ref{fig:dmix-tau}: Large open, filled and small open symbols correspond to models with $\feoh = -4$, $-3$, and $-2$, respectively. }  
Green and light blue colors represent the C22Ne models with the initial \nuc{14}{N} abundance $Y_{\rm N, ini} = 1.0$ and $0.1 Y_{\rm C, \odot}$, respectively.   
Black, orange, and red colors represent the R13C models with the mixing parameter $\dmix = 0.003$, 0.01, and 0.03, respectively. 
\red{ Symbols connected by lines represent models with the same metallicity of $\feoh = -3$ (shown as filled symbols) and the same mixing parameters, ordered by increasing maximum temperature. }
Thin black dotted lines on the top and middle panels are for C13C mode.
}
\label{fig:f-ntau-Ne} 
\end{figure}

During the recurrent shell flashes, the yield abundance of the elements \red{converges to a certain pattern by multiple neutron irradiations}.
 Figure~\ref{fig:f-ntau-Ne} shows the asymptotic efficiency factors $f_{\rm Z, asymp}$ for Sr, Ba and Pb, defined as \red{the asymptotic s-process abundance distributions} normalized by the pristine iron abundance.
The different models have different initial $Y_{\rm N, ini}$ and temperatures.
The transvers axis is the accumulative neutron exposure in the helium flash convection, defined as
\begin{eqnarray}
\ntauexpa = \sum_{i=1}^{n} \ntauexpt{}_{,n-i+1} \big ( \prod_{j=n-i+1}^{n} r_{j} \big) ,
\label{eq:nexpac} 
\end{eqnarray}  
where $r_i$ stands for the overlapping factor between the fraction of mass contained by the helium-flash convective zones during the $(i-1)$-th and $i$-th thermal pulses, and  the $\ntauexpt{}_{,i}$ is the total neutron exposure of the $i$-th shell flash (here we set $r_n = 1$).
  In the asymptotic limit, $\ntauexpa = \ntauexpt / (1-r)$ with the changes in the total neutron exposure during the recurrence ignored.
The asymptotic efficiency factors for the C22Ne mode, $f_{\rm Z, asymp}$, are much larger than those for the C13C mode because of the multiple neutron irradiations  \citep{ulrich73,iben75b,truran77}.
The dependence of $\ntauexpa$ on different models with different temperatures is not as sensitive as the amount of burnt \nuc{22}{Ne}. 
For example, the fraction of burnt \nuc{22}{Ne}, which is expected to be proportional to $\ntauexpa$, increases from 3.5\% to 55\% between Models M844 and M850, but the increment of $\ntauexpa$ is only a factor of 2.7.
This is because the amount of neutron poison \nuc{25}{Mg} also increases with the maximum temperature reached during the shell flash, which prevents the neutron exposure from becoming too large.  
For models with even higher temperatures, $\ntauexpa$ decreases.  
   Model M857 lowers $\ntauexpa$ by a factor of 6.7 as compared with Model M850 because of the larger production of \nuc{20}{Ne} and \nuc{24}{Mg}. 
Consequently, Model M850 attains the largest $\ntauexpa$. The C22Ne mode works effectively as the s-process nucleosynthesis only for the models between M847 and M853 with the narrow range of temperature between $T_{\rm He}^{\rm max} \simeq 3.0 \mhyph 3.4 \times 10^8$ K and the largest initial abundance of \nuc{14}{N}.

\subsection{Radiative \nuc{13}{C} Burning (R13C) Mode}

We study the repeated cycles of \nuc{13}{C} burning in the pocket and \nuc{22}{Ne} burning in the helium convection that occur during the shell flash.
In these cycles, the processed materials from the pocket are diluted by the helium convection and then further processed by the \nuc{22}{Ne} burning.
 After repeated shell flashes, the helium convection eventually leads to converged abundances of strontium, barium, and lead, as discussed in \S~\ref{sec:c22ne}. 

We set the pocket size to 1/20 of the stellar core, i.e., $\zeta_{\rm p} = 1/20$, and the temperature at $T = 10^8$ K.   
The amount of protons that mix into the pocket can vary depending on the physical conditions in the outer core and inner envelope.
This amount is expressed by the mixing parameter, $\dmix$, as in the C13C mode.
For the shell flashes, \red{we set the initial abundance of nitrogen to the solar abundance of carbon, $Y_{\rm N, ini} = Y_{\rm C,\odot}$}, which is the observed upper limit, in order to maximize the effect of the CNe22 burning.

The limited survey for the parameter range of the R13C mode can be justified for the following reasons.
First, the s-process nucleosynthesis by the R13C mode results in limited enrichment in the abundances of s-process elements.
The most efficient production of s-process elements can be achieved at $\log T \simeq 8.1$ according to our simulations.
In addition, the total amount of s-process elements in the envelope will be influenced by the pocket size, rather than the temperature in the pocket.
Therefore, changing the parameters for the pocket have much less significant impact on the abundances of s-process elements compared with the C13C mode.

In Model M840, \red{the asymptotic abundance distributions} are almost entirely made up of the materials from the pocket, which have been diluted by a factor of $\sim \zeta_p / (1-r) = 1/10$.
This is because \nuc{22}{Ne} burning has a relatively minor effect on the materials, except for the iron group elements.  
In Model M850, the \nuc{22}{Ne} burning is the main process that produces the first peak elements. The materials from the diluted pocket, on the other hand, are the main source of the second and third peak elements.

\begin{figure}

\includegraphics[width=\columnwidth,bb=0 0 380 300, clip]{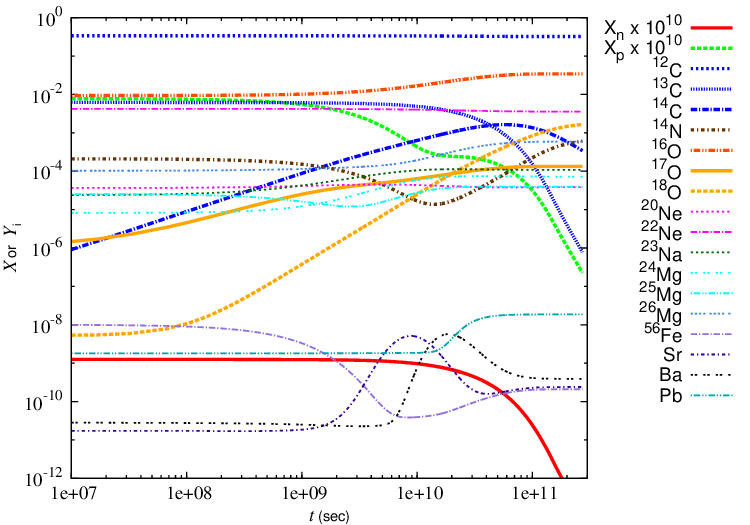}
\includegraphics[width=\columnwidth,bb=0 0 430 300, clip]{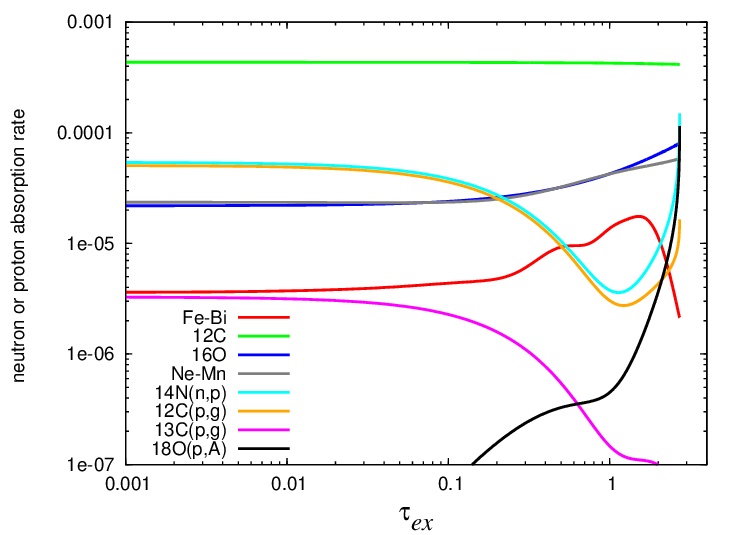}

\caption{
Progress of nucleosynthesis ({\sl upper panel}) and the variations in the neutron or proton absorption rates,  $\Sigma_{\rm A} \sigma_{\rm A}{}^{(n)}\rm{Y}_{\rm A}$ or $\Sigma_{\rm A} \sigma_{\rm A}{}^{(p)}\rm{Y}_{\rm A}  (\rm{Y}_{\rm p}/ \rm{Y}_{\rm n})$, ({\sl lower panel}) during the \nuc{13}{C} burning in the \nuc{13}{C} pocket for the model M840 with $\dmix = 0.03$ and $\feoh = -2$. 
Lines are the same as in Fig.~\ref{fig:22Ne-progress}, except that the proton absorption rate by \nuc{12}{C} is displayed in the lower panel. }
\label{fig:SYR13CP}
\end{figure}

Figure~\ref{fig:SYR13CP} shows the progress of nucleosynthesis and the neutron absorption rates for \red{the asymptotic abundance distributions} in the pocket.
The pocket also contains oxygen and Ne/Mg isotopes, which have accumulated during the recurrent shell flashes.
This means that the neutron poisons are much more abundant from the beginning. 
As seen in the bottom panel, even for $\feoh = -2$, oxygen and \red{the light elements from Ne to Mn} overwhelm heavy elements in the neutron absorption by an order of magnitude.
This is because \nuc{16}{O} increases more than the added \nuc{13}{C} due to the neutron recycling chain by \nuc{12}{C}.
Na and Mg isotopes also increase by neutron capture at the expense of \nuc{22}{Ne}, which has a smaller neutron absorption cross section.

As seen in the top panel, the proton density is much larger than the neutron density throughout the simulation. This is different from the C13C and C22Ne modes, where the neutron density is typically higher than the proton density. 
The initial abundance of \nuc{14}{N} is set to 0.032 times the \nuc{13}{C} abundance, the same as in \citet{gallino98}.  
The \nuc{14}{N} abundance decreases via the $(n, p)$ reaction, but later increases via the $\beta$-decay of \nuc{14}{C} beyond the initial abundance.  
As seen in the bottom panel, at the beginning, \nuc{14}{N} surpasses \nuc{16}{O} in the neutron absorption rate because of its much larger cross section. However, the protons released by \nuc{14}{N} are mostly captured by \nuc{12}{C} and eventually revert to neutrons.  
It is only in the later phase, after the abundance of \nuc{18}{O} grows larger than $Y_{\rm {}^{18}O} \simeq 4 \times 10^{-4} Y_{12}$  
that \nuc{18}{O} absorbs more protons than \nuc{12}{C} to remove neutrons.  
However, \nuc{14}{N} has little impact on the neutron density since \nuc{13}{C} has burnt appreciably by the time when the proton capture of \nuc{18}{O} catches up with the neutron absorption by \red{the light elements from Ne to Mg}.  
\nuc{14}{N} may play a dominant role as the neutron poison if the initial abundance of \nuc{14}{N} is larger than that of \nuc{13}{C} to produce more \nuc{18}{O} by more than an order of magnitude (c.f., \cite{gallino98,gallino10}).   

As shown in Figure~\ref{fig:dmix-tau}, the neutron exposure achieved in the pocket, $\ntauexptp$, is significantly lower than that of the C13C mode, especially for models with higher temperatures.  
For example, $\ntauexpt$ decreases from 2.8 to $1.05 \hbox{ mb}^{-1}$ between Models M840 and M850, while it can be as large as $\simeq 6 \hbox{ mb}^{-1}$ for the C13C mode at $\dmix = 0.03$.  
   In the pocket, the neutron density remains below $n_n = 8.3 \times 10^7 (7.6 \times 10^8) \hbox{ cm}^{-3}$ for $\dmix = 0.003 (0.03)$, during the inter-flash phase of $7.6 \times 10^3$ yr. 
In contrast, the C13C mode can produce neutron densities of $n_n = 0.55 \mhyph 3.2 \times 10^{11} (0.16 \mhyph 1.7 \times 10^{12}) \hbox{ cm}^{-3}$ for $\dmix = 0.003(0.03)$ in Model M840 with hydrogen mixing over $10^8$ sec.

\red{
There are two important factors affecting the total neutron exposure, $\ntauexptp$, in the 13C pocket; the one is $\dmix$ and the other is temperature in the shell flash.  The larger the $\dmix$ is, the larger the $\ntauexptp$ is.  
And higher temperature in the shell flash results in smaller $\ntauexptp$ in 13C pocket, because the amount of the accumulated neutron poisons ( Ne-Mg isotopes) increases in the pocket. These effects are clearly seen in Fig.~\ref{fig:dmix-tau}.
}

The asymptotic efficiency factors, $f_{\rm Z,asym}$ , of the R13C mode (black, orange and red symbols) are plotted in Fig.~\ref{fig:f-ntau-Ne} as a function of $\ntauexpa$, which is defined in the same way as eq.~(\ref{eq:nexpac}). 
The processed materials in the pocket are determined by the mixing parameter, as in the C13C mode.
The contribution from the C22Ne mode grows larger for the models with higher temperatures.  
We see that $f_{\rm Sr,asym}$ tends to be separated in inversely proportional to the metallicity, and $f_{\rm Ba,asym}$ to a lesser extent, for Models M840 and M844 of the R13C mode with $\dmix = 0.03$.
This means that the Ne/Mg seed phase takes place for these models.
In this case, the phase transition from iron to Ne/Mg seed phases occurs at larger $f$-values, as compared with the C13C mode (thin black dotted line), in particular for low metallicity.
Consequently, the R13C mode covers a narrower range of $f$-values during the s-process nucleosynthesis than the C13C mode, since the upper bound is limited by the pocket size and the lower bound by the transition point from iron to Ne/Mg seed phases (see also Fig.~\ref{fig:correlation_f}).




\section{Correlations among Sr, Ba and Pb}\label{sec:correlations}

\begin{figure}
\includegraphics[width=\columnwidth,bb=0 0 1053 1053, clip]{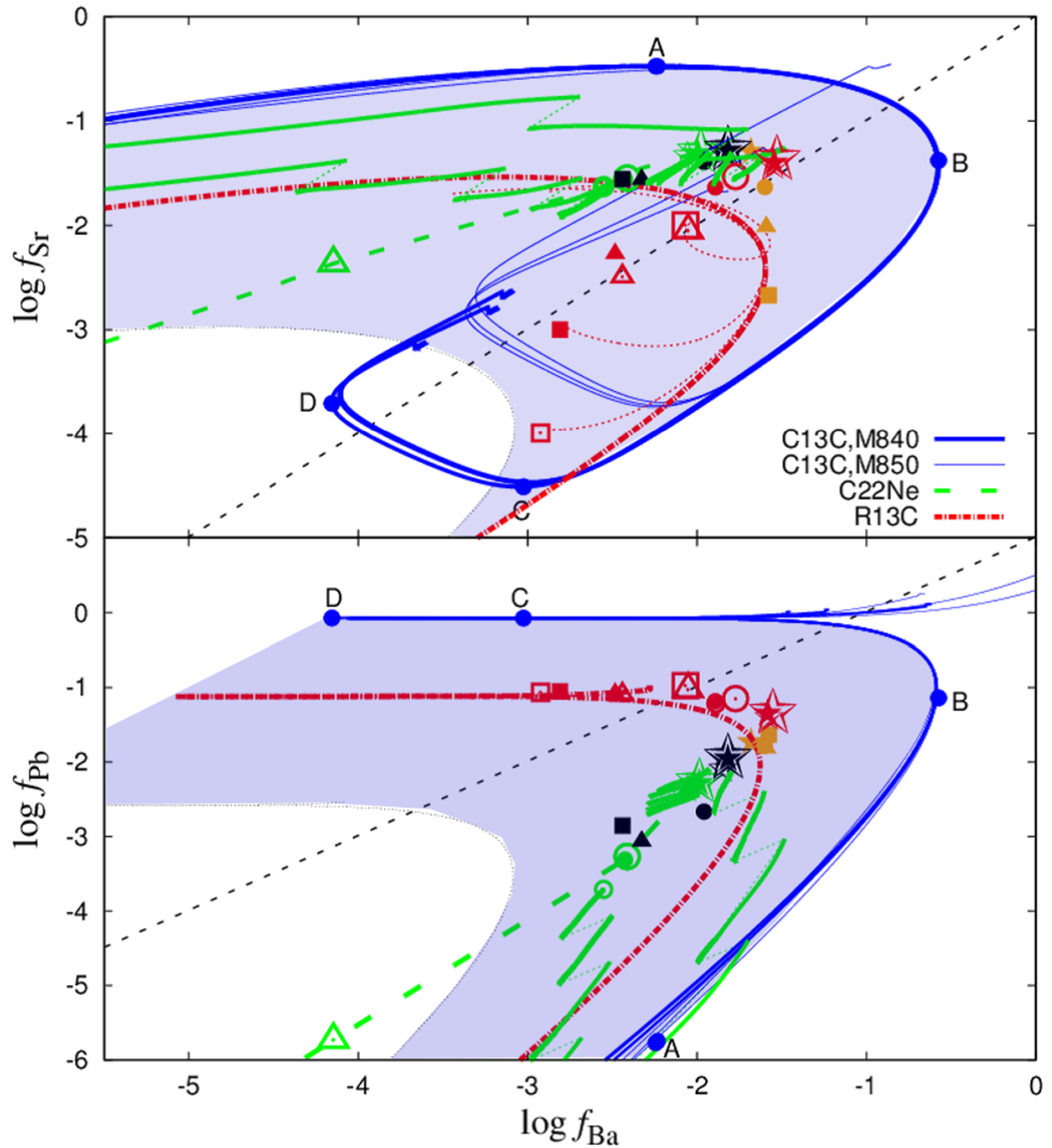}

\caption{
 The correlation diagrams of the efficiency factors of Ba and Sr (upper panel) and of Ba and Pb (lower panel) for the three modes of \sps\ nucleosynthesis.   
   The trajectories of \reded{materials} of the C13C mode are drawn as a function of the mixing parameter up to $\dmix = 0.03$ for Models M840 and M850 (thick and thin blue lines) with the metallicity of $\feoh = -4$, -3, and -2.
The f-values become larger for smaller metallicity on the Ne/Mg seed phase.
   Points A and B denote the stages of maximum $f_{\rm Sr}$ and $f_{\rm Ba}$ and points C and D do the transitions to the Ne/Mg seed phase for Sr and Ba, respectively, for Model M840 with $\feoh = -3$. 
Shaded (blue) areas denote the materials of the C13C mode, diluted by the carbon later dredged up in the envelope via TDU during the shell flashes of C22Ne mode (see text).
For the C22Ne mode, \red{the asymptotic abundance distributions} with $Y_{\rm N, ini} = Y_{\rm C, \odot}$ are plotted (green symbols with dashed line), and \red{the convergence of the abundances} during the recurrence are illustrated for Models M847 and M850 (thick green lines connected with dotted lines). 
   For the R13C mode, \red{the asymptotic abundance distributions} with the pocket size, $\zeta_P = 1/20$, and the mixing parameter, $\dmix = 0.003$, $0.01$, and $0.03$ are plotted (black, orange, and red symbols, respectively) and compared with the trajectory of the C13C mode, reduced by a factor of $\zeta_p / (1 - r) = 1/10$ (red dash-dotted curves).   
   Black broken line denotes the steady-state abundance ratios, $f_{\rm Sr} / f_{\rm Ba} =  \sigma_{n, \rm Ba} / \sigma_{n, \rm Sr}$. 
The symbols are the same as those in Fig.~\ref{fig:dmix-tau}.
}
\label{fig:correlation_f}
\end{figure}

We have discussed the properties of the three modes of the three s-process peak elements.
Their characteristics can be more easily understood by looking at the correlations between the efficiency factors of the elements.  
   Figure~\ref{fig:correlation_f} compares the behaviors of the three modes on the correlation diagrams of $f_{\rm Ba}$ vs. $f_{\rm Sr}$ and $f_{\rm Ba}$ vs. $f_{\rm Pb}$.  
For the C13C mode, we plot the trajectory of the s-process nucleosynthesis as a function of the neutron exposure for different mixing degrees, $\dmix$, taken from Models M840 and M850.
As for the C22Ne and R13C modes, the asymptotic efficiencies are taken from various models with the initial \nuc{14}{N} abundance, $Y_{\rm 14N, ini} = Y_{\rm C}{}_\odot$, the overlapping factor, $r=0.5$, and the pocket size, $\zeta_p = 1/20$.  
Since the progress of the s-process efficiency is independent of the metallicity in EMP stars with $\abra{Fe}{H}\lesssim -2$, as shown above, we can compare the results for the different metallicity on the same diagrams.  

The C13C mode provides a wide range of $f$-values, and the trajectories of the $f$-values move clockwise in the Ba-Sr diagram and anticlockwise in the Ba-Pb diagram as the neutron exposure increases.  
   In the Ba-Sr diagram, $f_{\rm Sr}$ and $f_{\rm Ba}$ first increase to reach local maxima (points A and B) and then decrease by by $3 \sim 4$ dex to hit minima (points C and D) and then increase again.
The transition from the Fe seed phase to the Ne/Mg-seed phase occurs earlier at larger $f$-values in the models of higher temperatures due to the larger production of Mg seeds during the shell flashes.  
On the Fe seed phase, the $f$-values follow almost the same trajectories, regardless of the models and the metallicity.
Along the Fe seed phase, the Ba/Sr ratio increases beyond the steady-state value (black dashed line) of $f_{\rm Ba}/ f_{\rm Sr} = \sigma_{n, \rm Sr} / \sigma_{n, \rm Ba}$ or $\abra{Ba}{Sr} = 0.88$ and rises up to $\abra{Ba}{Sr} \simeq 2.5$. 
In contrast, on the Ne/Mg seed phase, the $f$-values branch off in inverse proportion to the metallicity.
The smaller the metallicity, the larger the $f$-value, as shown in Fig.~\ref{fig:tauC13C}.
  The $\abra{Ba}{Sr}$ ratio on the Ne/Mg seed phase decreases across the steady-state value down to $\abra{Ba}{Sr} \simeq -0.8$, turning to increase and gradually approaching to the steady-state value.

   In the Ba-Pb diagram, on the other hand, $f_{\rm Pb}$ keeps increasing through the Fe and Ne/Mg seed phases.
As a result, the trajectory of $f_{\rm Pb}$ move upwards as the mixing parameter increases.  
After hitting the maximum of $f_{\rm Ba}$ (point B), the trajectory shifts leftwards along the trajectories of $f_{\rm Pb} \simeq 1.0$ and then turns back rightwards after hitting the minimum of $f_{\rm Ba}$ (point D).  
Meanwhile, the Pb/Ba ratio increases greatly up to $\abra{Pb}{Ba} = 4.6$, which is much higher than the steady-state value of $\abra{Pb}{Ba} = 1.46$.   
On the Ne/Mg branch, the Pb/Ba ratio decreases across the steady-state value, and then turns to increase and approach the steady-state value.  

In contrast, the path of the processed materials of the C22Ne mode runs in the middle of the domain, surrounded by the path of the C13C mode.  
The neutron exposure is kept low even in the optimal models of M850 due to the production of strong neutron poison, \nuc{25}{Mg}. 
We depict \red{the convergence of the abundances} during the recurrent shell flashes (thick green lines connected with dotted lines).
   Following \citet{ulrich73}, we can write the accumulated efficiency, $f_{\rm Z, n}$, in the $n$-th shell flash as the superposition of the $f$-value of the single irradiation, \blue{$f_{\rm Z, s}$}:
\begin{eqnarray}
f_{\rm Z, n} (\ntauexpt) &=& (1-r) \sum_{k=1}^{n} r^{k-1} f_{\rm Z, s} (k \ntauexpt) \nonumber \\
   &+& r^{n-1} f_{\rm Z, s} (n \ntauexpt). 
\label{eq:f-n-ne22}
\end{eqnarray}
   For small neutron exposure, $f_{\rm Z, s}$ increases steeply as a function of neutron exposure.  
 the accumulated efficiency is dominated by the last term in the right-hand side, where the seed isotopes are $n$ times irradiated by neutrons.
The accumulated efficiency has its maximum on the way to \red{the asymptotic abundance distribution}, as seen in the thick green lines connected with dotted lines in Fig.\ref{fig:correlation_f}. 
When the $f_{\rm Z, n}$ goes over the maximum, the last term no longer exceeds the other terms with seeds irradiated less than $n$ times.
 In fact, such a maximum occurring in the C22Ne mode has been shown by \citet{truran77}.
The abundance ratios of Ba to Sr elements increase with $\ntauexpt$, but are smaller than that of the C13C mode and can never exceed the ratios set by the maxima of $f_{\rm Z, n}$.

So far, we have considered the C13C and C22Ne modes separately. 
However, in reality, after the He-FDDM enriches the surface carbon abundance beyond $\abra{CNO}{H}\simeq -2.5$,
  the C22Ne mode enriches surface carbon abundance via TDUs up to $\abra{C}{H}\simeq 0$ at which
 the envelope mass is stripped by the carbon dust-driven wind (see \S~\ref{sec:c13c}).   
For the carbon abundance, we need to take into account the processed materials of both the C13C and the succeeding C22Ne modes. 
The averaged efficiency factor, $f_{\rm Z, av}$, combining the C13C with the C22Ne modes, is effectively reduced by the increment, $\Delta \abra{C(+N)}{H}$, of the carbon abundance via TDUs during the later shell flashes without the hydrogen engulfment.
This can be expressed as
\begin{equation} 
\log f_{\rm Z, av} = \log f_{\rm Z} - \DeltaCNoH.  
\label{eq:DeltaCNoH}
\end{equation}
   The range of carbon increment is given by
\begin{equation} 
0 \le \DeltaCNoH \lesssim 2.5,
\end{equation} 
This is because TDUs can enrich the abundances in the envelope from $\abra{CNO}{H} \simeq -2.5$ to $\abra{C}{H} \simeq 0$.
\red{ Shaded areas in Fig.~8 denote the range of $f_{\rm Z, av}$. }

\begin{figure}
\includegraphics[width=\columnwidth,bb= 0 0 415 307, clip]{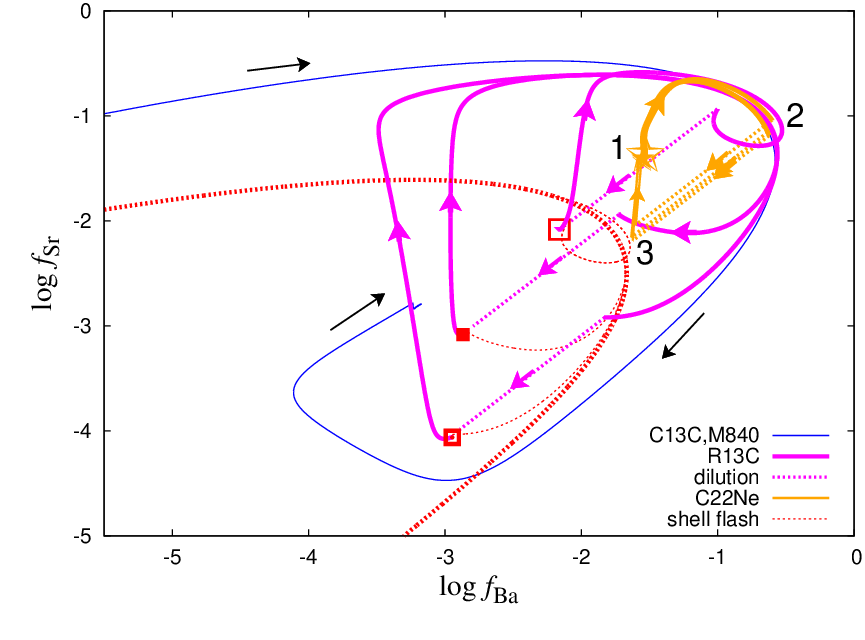}

\caption{
   The asymptotic behaviors of the R13C mode on the same diagram of the efficiency factors of Ba and Sr as in Fig.~\ref{fig:correlation_f} for Models, M840 (magenta lines) and M850 (orange lines), with $\dmix = 0.03$.  
   Thick solid curves starting from each symbol \red{along arrows} ( 1 $\rightarrow$ 2 ) describe the trajectories of \sps\ nucleosynthesis by the radiative \nuc{13}{C} burning in the pocket.  
   Then, dotted lines (2 $\rightarrow$ 3) denote the dilution when the \sps\ materials in the pocket are mixed in the helium flash convection during the succeeding shell flash.  
   For Model M850, vertical solid lines up to each symbol (3 $\rightarrow$ 1) denote the addition of the materials of the C22Ne mode, mainly composed of the 1st peak elements.
}
\label{fig:correlation_f_R13C}
\end{figure}

For the R13C mode, the nucleosynthesis in the \nuc{13}{C} pocket proceeds similarly to the C13C mode, but with a different timescale.
The $f$-values vary as a function of the mixing parameter $\dmix$.   
The materials synthesized in the pocket are involved in the helium shell flash (C22Ne mode), which is repeated.
Similarly to the C22Ne mode in eq.~(\ref{eq:f-n-ne22}), the contribution of the \sps\ nucleosynthesis in the pocket to the asymptotic efficiency during the recurrence, is written in the form;
\begin{eqnarray}
   f_{\rm Z, asymp} = \frac{1-r}{1-(r-\zeta_p)} \sum_{k=1} \left[ \frac{\zeta_p}{1-(r-\zeta_p)} \right]^{k} f_{{\rm Z, s}} (k \ntauexptp)  
\label{eq:f-R13C}
\end{eqnarray}
where $\ntauexptp$ is the neutron exposure in the pocket.
For small $\ntauexptp$, the asymptotic efficiency tends to be dominated by the effect of multiple irradiation, as in the case of the C22Ne mode.
For larger  $\ntauexptp$, on the contrary, the asymptotic efficiency tends to be dominated by the materials of the final shell flashes, i.e., the first term in eq.~(\ref{eq:f-R13C}).
In reality, in Fig.~\ref{fig:correlation_f}, we see that the models with $\dmix = 0.003$ (black symbols)  fall along the trajectory of the asymptotic efficiencies of the C22Ne mode.
For models with $\dmix = 0.03$ (red symbols), the contribution of the nucleosynthesis in the pocket is given by the trajectory of the C13C mode \red{reduced by a factor of $\zeta_{p} / (1-r)$} (red dash-dotted line).
 
Figure~\ref{fig:correlation_f_R13C} illustrates the progress of nucleosynthesis of the R13C mode for Models M840 (magenta lines) and M850 (orange line), with the pocket size $\zeta_p =1/20$ and the mixing parameter $\dmix=0.03$.   
   In the pocket, the $f$-values of both models approach the trajectories of the C13C mode, despite large differences in the initial abundances (magenta and orange thick solid curves).   
Model M840 enters the Ne/Mg seed phase much earlier at much larger $f$-values than the C13C mode, which also depends on the metallicity.
Model M850, on the contrary, the s-process nucleosynthesis terminates on the Fe seed phase due to the accumulation of Ne/Mg isotopes as neutron poisons.  
During the succeeding shell flash, the processed materials from the pocket are diluted in the flash convection (dotted lines). For Model M850, the 1st peak element processed by the \nuc{22}{Ne} burning is added (blue vertical thick solid line from 3 to 1).  
The abundances of s-process elements are only slightly modified by the R13C mode, except for the first peak elements.
This is because the neutron exposure from the C22Ne mode is much smaller than in the pocket.
The asymptotic efficiency of the s-process in the pocket is given by $f_{\rm Z, asymp} \simeq [\zeta_p / (1-r)] f_{\rm Z, s} (\ntauexptp)$ from eq.~(\ref{eq:f-R13C}) for $\zeta_p \ll 1$.

The asymptotic efficiency of the s-process in the pocket increases as the pocket size increases, but there is a maximum efficiency that is reached when $\zeta_p = r$.
This results in \red{the asymptotic abundance distributions} of
\begin{eqnarray} 
   f_{\rm Z, asymp} = (1-r) r \sum_{k=1} r^{k-1} f_{{\rm Z, s}} (k \ntauexptp)  \\  
\label{eq:f-R13C_max}
\end{eqnarray} 
   The largest efficiencies are likely to be much smaller than the maxima achieved by the C13C mode, because $(1-r) r$ is always less than or equal to $1/4$ and because $f_{{\rm Z, s}} (\ntauexp)$ decreases sharply for larger $\ntauexp{}_{\rm , t}$, except the third peak element.

Consequently, the R13C mode can only cover a limited range of $f$-values and the abundance ratios in comparison with the C13C mode.     
For small \blue{amount of mixing} in the pocket, the abundance ratios of s-process elements by the R13C mode fall along the same line as the C22Ne mode.
For large \blue{amount of mixing}, the ratios lie along the trajectory of the C13C mode \red{reduced by a factor of $\zeta_{p} / (1-r)$}.
The maxima reached by the trajectory of the R13C mode is four times smaller than that of the C13C mode, but the minima are much larger due to the accumulation of Ne/Mg isotopes.

\section{Characteristic and efficiency of the \lowercase{$s$}-process nucleosynthesis}

In this section, we discuss the role of the s-process nucleosynthesis in CEMP stars by comparing the s-process materials processed in EMP-AGB stars with the observed abundances of neutron-capture elements in CEMP stars.
This offers the prospect of finding the actual conditions that the s-process nucleosynthesis works and also provides the basis for inquiring into the origins of the subgroups of CEMP stars.

\begin{figure}
\includegraphics[width=\columnwidth,bb= 0 0 917 954, clip]{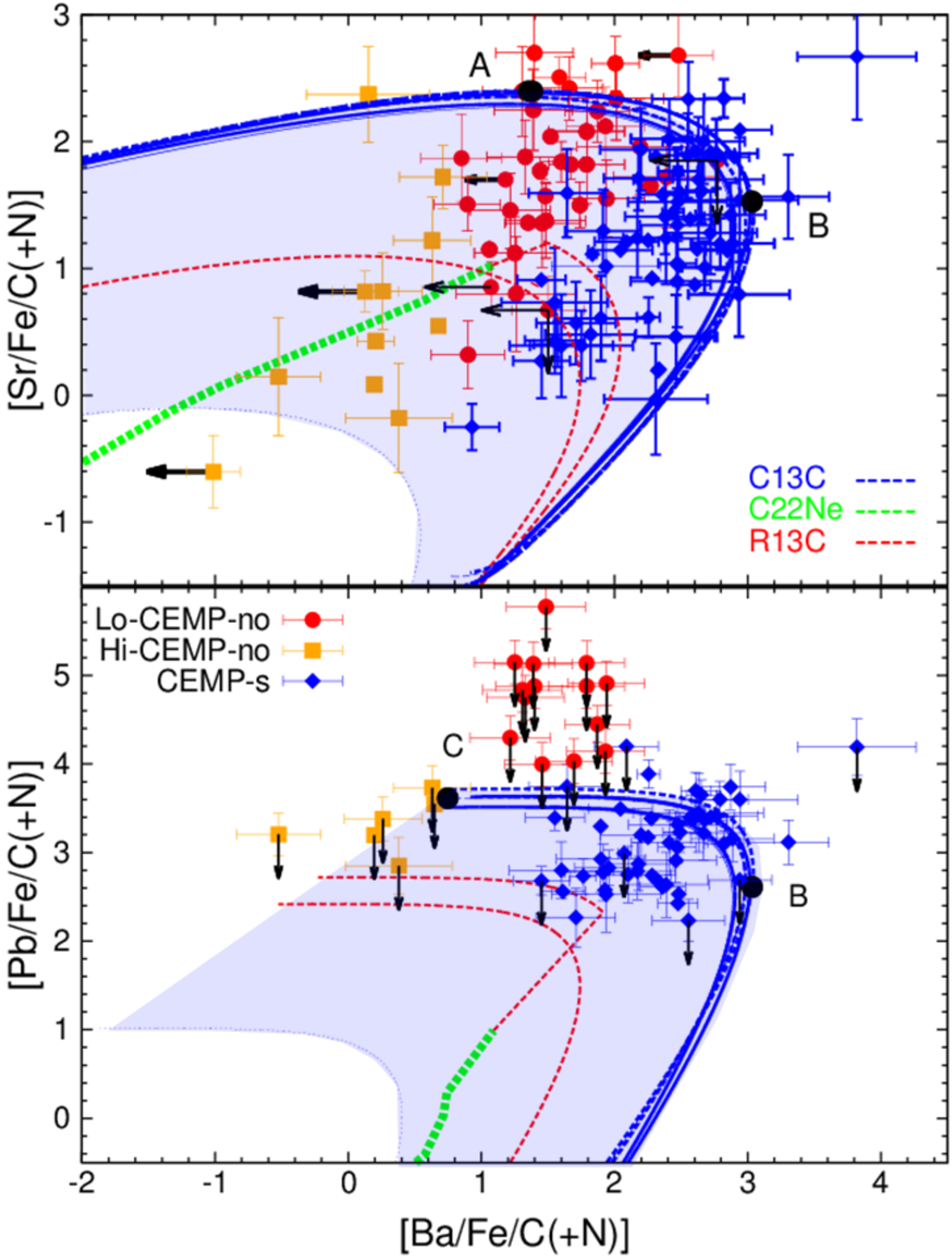}
\caption{ 
   Correlation diagrams of Ba and Sr (upper panel) and Ba and Pb (lower panel) for the predictions of the s-process nucleosynthesis, compared with the observed data of CEMP stars. 
The enrichments of Sr, Ba, and Pb, normalized by the carbon abundance, are taken as the axes (see eqs.~[\ref{eq:abrancnZ}] ). 
   Solid and broken blue curves denote the trajectory of the materials of the C13C mode with different degrees of hydrogen engulfment, taken from Models M840 and M844, respectively (see Table~\ref{tab:param}).  
Points A and B denote the locations of the maximum $\abranc{Sr}$ and $\abranc{Ba}$ for Model 840 with $\feoh = -3$.  
Blue shaded area denotes the trajectory of the C13C mode, diluted by carbon dredged up during the succeeding shell flashes up to the maximum increment of carbon abundance, $\DeltaCNoH = 2.5$ (see text).  
   Thick green broken lines denote \red{the asymptotic abundance distributions} of the C22Ne mode for the models of various masses with the initial nitrogen abundance, $Y_{\rm N,}{}_i = Y_{\rm C, \odot}$, and the overlapping fraction of the helium flash convection, $r = 0.5$, where the largest materials are given by Model M850.  
   The materials of the R13C mode are approximated by the trajectory of the C13C mode, reduced by $\zeta_p/(1-r) ( = 0.1)$ (the materials of single shell flash) and by $\zeta_p (1-r) / (1-r+\zeta_p)^2$ (\red{the converging abundances} after the recurrence), for the pocket size and the overlapping fraction, $\zeta_p = 1/20$ and $r = 0.5$ (red dotted lines).    
 The observed data of CEMP-s (blue diamonds), Lo-CEMP-no (red circles), and Hi-CEMP-no (brown squares) stars are plotted. 
}
\label{fig:BaCSrC2} 
\end{figure}

\subsection{Dilution free indicator of $ \abranc{Z}$}

We introduce new correlation diagrams among Sr, Ba, and Pb in Figure~\ref{fig:BaCSrC2}.
The correlation between Ba and Sr (upper panel) and between Ba and Pb (lower panel) provide comparisons of the materials of the s-process nucleosynthesis with the observations of CEMP stars.
The abundance data are taken from the SAGA database (June 18, 2015 version; Suda et al. 2008).
The coordinate axes for element Z are normalized by carbon abundances as defined by 
\begin{eqnarray}
  \abranc{Z} \equiv \abra{Z}{H} - \abra{Fe}{H}_{\rm p} - \abra{C}{H} 
\label{eq:abrancnZ}
\end{eqnarray} 
together with the normalization by the pristine iron abundance, $\abra{Fe}{H}_{\rm p}$.  
The advantage of using eq.~(\ref{eq:abrancnZ}) is that the s-process materials and carbon are carried together all the way from the materials dredged-up into the envelope of the primary AGB stars until wind accreted on to the secondary and now observed as CEMP stars.
The abundance ratios of the s-process and the carbon in the dredged-up materials by the helium flash convection are unchanged through the whole mass transfer process, because the s-process and carbon abundances in the envelope are much smaller than that in the dredged-up materials.
Therefore, $\abra{Z}{C}$ is free from suffering from dilution effect, different from $\abra{Z}{H}$ and $\abra{Z}{Fe}$.
 \red{As a result, $\abranc{Z}$ is also free from the dilution effect.}

For CEMP stars, \red{the quantity $\abranc{Z}$ can be readily computed as}
\begin{eqnarray}
   \abranc{Z} =  \abra{Z}{Fe} - \abra{C}{H}, 
\label{eq:abrancnZ2}
\end{eqnarray} 
because the Fe abundance in the dredged-up materials is much smaller than that in the envelope. This allows us to directly compare the observed abundances and the predicted s-process materials, which were synthesized in the helium flash convective zones.

\subsection{Comparison with the Observed Abundances}

The use of the carbon normalized enrichment, $\abranc{Z}$, allows us to directly compare the observed abundances and the theoretical predictions for CEMP stars, regardless of the metallicity.  
We define the carbon abundance of CEMP stars as $\abra{C (+ N)}{Fe} \equiv \log [ (Y_{\rm C} + Y_{\rm N}) / Y_{\rm C, \odot}  ] -\abra{Fe}{H} (\ge 0.7)$.
Here, $Y_{\rm C}$ and $Y_{\rm C, \odot}$ are the mole abundances of carbon and solar carbon abundances, respectively, with the nitrogen abundance, $Y_{\rm N}$, added if available.
This definition takes into account the possible processing of dredged-up carbon by CN cycle reactions during the He-FDDM and in the envelope of AGB stars \citep{suda04,aoki07}.
In Fig.~\ref{fig:BaCSrC2}, we plot CEMP stars with $\abra{Fe}{H}<-2$.
   
Here, we divide CEMP-no stars (those with $\abra{Ba}{Fe} < 0$) into two subgroups based on the value of $\abrancn{Ba} = 0.8$, which corresponds to the lowest limit of CEMP-s stars.
We define Lo-CEMP-no stars as those with $\abrancn{Ba} \ge 0.8$ and Hi-CEMP-no stars as those with $\abrancn{Ba} < 0.8$.
The carbon abundances in Lo-CEMP-no stars are in the same range as CEMP-s stars, while those in Hi-CEMP-no stars are smaller than CEMP-s stars.
This means that the C22Ne mode must operate more in Hi-CEMP-no stars than in CEMP-s stars in order to lower $\abrancn{Ba}$ to below 0.8.
This difference in the C22Ne mode operation also results in a difference in the carbon abundances between the two subgroups.
   Lo-CEMP-no stars mostly have $\abra{C(+N)}{H} \simeq -1.5$, while Hi-CEMP-no stars have $-1.3 \lesssim \abra{C(+N)}{H} \lesssim 0$, which is similar to the range of carbon abundances in CEMP-s stars.

The Sr/Ba correlation diagram (upper panel of Fig.~\ref{fig:BaCSrC2}) shows that all CEMP stars, including CEMP-s and CEMP-no stars, are located in the domain encompassed by the trajectories of the C13C mode with the pristine iron as seed and those in combination with the C22Ne mode (blue shaded area).
Some CEMP stars lie along the trajectory of the C22Ne mode (green broken line), \red{which is expected because this mode produces more Sr than Ba.}
However, there are also a few CEMP-s stars that lie outside the trajectories of the C13C mode.
One exception is SDSS1707+58, which has the largest enrichments of Ba and Sr among CEMP-s stars. This star also has large enhancements of Na and Mg \footnote{  
   SDSS1707+58 with $\abra{C}{Fe} = 2.1$ at $\feoh = -2.52$ marks the largest enrichments of Ba and Sr ($\abra{Ba}{Fe} = 3.40$ and $\abra{Sr}{Fe} = 2.25$) among CEMP-s stars, and shows the enrichments of Na and Mg, $\abra{Na}{Fe} = 2.71$ and $\abra{Mg}{Fe} = 1.13$ \citep{aoki08}.  \hfill\eject\indent
   On the other hand, CS29528-028 with $\abranc{Ba} = 3.36$ \citep{aoki07} and SDSSJ1836+6317 with $\abranc{Ba} = 3.2$ \citep{aoki13} do not appear in this diagram because of the absence of the Sr abundances; 
   they also lack the N abundances.
}, which suggests that it received a large amount of material from the AGB companion star.
Several Lo-CEMP-no stars lie slightly above the trajectories of the C13C mode. This is likely due to the fact that these stars have larger pristine abundances of Sr and carbon.
\red{It is to be noted that carbon-normal stars with metallicities lower than [Fe/H] $< -2$ have $\abranc{Sr}\simeq 2.56$ on average, which is higher than most of CEMP stars.}
In general, the distribution of CEMP stars in the Sr/Ba correlation diagram is continuous, with no apparent gaps between the three groups of stars.
This suggests that all CEMP stars may have formed in a similar way, with the main difference being the amount of material that was accreted from the AGB companion star.

CEMP-s stars are found along the trajectory of the C13C mode near the maximum of $\abrancn{Ba}$ and in the domain below them.
This is consistent with the idea that these stars formed from material that was processed by the C13C mode, followed by the carbon dredge-up by the C22Ne mode. 
Hi-CEMP-no stars with $\abrancn{Ba} < 0.8$ have a wide range of $\abrancn{Sr}$ values.
Some of them lie along the trajectory of the C22Ne mode, while others lie in the domain of the C13C mode where they have smaller $\abrancn{Ba}$ values.
   Lo-CEMP-no stars occupy a wedge-shaped domain between CEMP-s and Hi-CEMP-no stars.
   They have [Sr/Ba] ratios that are larger than those of CEMP-s stars, but comparable to those of Hi-CEMP-no stars. One Hi-CEMP-no star has the largest [Sr/Ba] ratio, but it is still located on the trajectory of the C13C mode.
   

In this diagram, the R13C mode can only produce s-process materials that are compatible with the observed abundances in a narrow range of the size of the \nuc{13}{C} pocket.
This means that the size of the \nuc{13}{C} pocket would need to \red{increase} by two orders of magnitude in order to cover the entire range of observed abundances, \red{which is not realistic}.  
Theoretically there is an upper limit on the size of the \nuc{13}{C} pocket that can produce the largest efficiencies. 
This limit is only 1/4 times the maximum size that can be achieved by the the C13C mode (see \S4). 
\red{It is also argued by \citet{bisterzo10} that the maximum size of the \nuc{13}{C} pocket may be the twice the standard value because mixed protons are consumed by \nuc{13}{C} and form \nuc{14}{N} in the pocket.}

\red{
\citet{abate15a} and \citet{bisterzo12} performed s-process nucleosynthesis simulations with AGB models that only include the R13C and C22Ne modes.
These models were able to reproduce the abundance patterns of many CEMP-s stars. However, the carbon abundances in these models were significantly higher than the observed values.
If the carbon abundance is taken into account in the fit to the observations, the fit may be worse.
}


In the Ba-Pb correlation diagram (lower panel), \red{the observed abundances of CEMP-s stars are consistent with the C13C mode on the Fe seed phase in combination with the C22Ne mode.
These two processes involve the production of carbon in the helium-burning shell, which is later dredged-up by the TDU.}
   For CEMP-no stars, only upper limits are available, so no meaningful constraints can be drawn. 

Consequently, the abundance variations of neutron-capture elements in observed CEMP stars are consistent with the materials predicted to be produced by the C13C and C22Ne modes, and their combination.  
The abundances of CEMP-s stars are consistent with those produced by the C13C mode, with the additional enhancement of carbon produced by the C22Ne mode in low-mass AGB stars. 
Hi-CEMP-no stars are attributed to the C22Ne mode in massive AGB stars, or to the C13C mode with greater dilution by the later dredged-up carbon, and/or with smaller hydrogen engulfment resulting in larger Sr/Ba ratios than CEMP-s stars.  
Lo-CEMP-no stars are continuously connected to CEMP-s stars with larger Sr/Ba ratios, and to Hi-CEMP-no stars. 
This contiguous arrangement is indicative of the common formation mechanism for all of these subclasses.  

These results support the hypothesis that binary mass transfer is responsible for the carbon enhancements in CEMP stars, which were born as the low-mass companions to AGB stars. The amount of carbon that they accrete depends on the binary separation, the efficiency of the s-process nucleosynthesis in the primary AGB star, and the degree of hydrogen engulfment.

\section{Conclusions and discussion}

  We have analyzed the observed properties of neutron-capture elements in carbon-enhanced metal-poor (CEMP) stars to explore the possibility that they formed through binary mass transfer.
To do this, we investigated the progress of the s-process nucleosynthesis in extremely metal-poor (EMP) AGB stars using a semi-analytical model of shell flashes.
We revealed the characteristics and distinctions of the three modes of s-process nucleosynthesis that have been proposed to date.
Our main results are summarized as follows:

\vskip 0.5pc\noindent 
{\bf (I) The $s$-Process Nucleosynthesis in EMP-AGB Stars}

A critical aspect of \sps\ nucleosynthesis in EMP stars with $\abra{Fe}{H} \lesssim -2$ is that the efficiency factors, defined by eq.~(\ref{eq:efffactor}), of ratio of the  mole abundances of \sps\ to pristine iron relatively insensitive to metallicity.
This is because iron nuclei and heavy elements no longer act as a neutron \red{absorber} in the low metallicity environment of $\abra{Fe}{H}<-2$.
Instead \nuc{16}{O} and its progenies, produced in the helium convective zone, act as neutron poisons and determine the neutron density and the neutron exposure.
\red{In other words, the number of heavy elements produced per iron is independent of metallicity during the Fe seed phase, as shown in Fig.~\ref{fig:tauC13C}.}
 The basic characteristics and distinctions of the three modes of \sps\ nucleosynthesis are as follows:

\vskip 0.5pc \noindent 
{\sl (i) \ the convective \nuc{13}{C} burning (C13C) mode:} 
   
The C13C mode is a neutron-capture process that is triggered by the hydrogen engulfment into the helium flash convective zone during an early stage of TP-AGB evolution in stars with $M \lesssim 3.5 \msun$ and $\feoh \lesssim -2.5$.
The processed materials are dredged-up to the surface by the helium-flash driven deep-mixing (He-FDDM).  
This mode is divided into two phases according to the degree of hydrogen mixing or neutron exposure.
During the first phase (Fe seed phase), the pristine iron-group elements act as seeds for the production of \sps\ elements.
In this phase, the $s$-process efficiencies of Sr and Ba, $f_{\rm Sr}$ and $f_{\rm Ba}$, first increase up to the maxima of $\log f_{\rm Sr} = 0.36$ and $\log f_{\rm Ba} = 0.26$, respectively, and then, decreases by three or four orders of magnitude with increasing neutron exposure.
When the efficiency factor hits the bottom and starts to increase again, the second phase (Ne/Mg seed phase) begins.
During this phase, Ne and Mg isotopes act as seeds for the production of \sps\ elements, triggered by the $\alpha$-captures of \nuc{14}{N} and \nuc{16}{O} in the helium zone.
   For the third peak elements, the $s$-process efficiencies of Pb, $f_{\rm Pb}$, keeps increasing up to $f_{\rm Pb}= 1.0$ with increasing neutron exposure. 

\vskip 0.5pc \noindent 
{\sl (ii) \ the convective \nuc{22}{Ne} burning (C22Ne) mode:} 

\red{This mode uses \nuc{14}{N} in the ash of hydrogen shell burning to synthesize \nuc{22}{Ne}, depending on the enhancement of surface carbon that is enriched by TDUs.}
The burning of \nuc{22}{Ne} and the s-process efficiencies increase with temperature, reaching a maximum for the model with $\log T_{\rm He}^{\rm max} = 8.50$ (M850).  
For higher temperatures, the production of neutron poisons of Ne/Mg isotopes prevents s-process nucleosynthesis, resulting in a sharp decrease in the s-process efficiencies. 
Therefore, the C22Ne mode is only effective in producing s-process materials in a narrow range of temperatures, $\log T_{\rm He}^{\rm max} \simeq 8.47 \mhyph 8.53$, which are reached during the shell flashes. 
This limits the site of the C22Ne mode to massive AGB stars with $M > 3.5 \msun$.
Because of the strong neutron poisons, the efficiencies are suppressed as low as $f_{\rm Ba} \simeq 0.01$, which is much smaller than the C13C mode, even in the optimal model of M850.

\vskip 0.5pc \noindent 
{\sl (iii) \ the radiative \nuc{13}{C} burning (R13C) mode} 

The R13C mode is a neutron-capture process that occurs in low-mass AGB stars, and is also known as the \nuc{13}{C} pocket hypothesis.
It is assumed that hydrogen is injected into the carbon-rich radiative helium zone, forming a \nuc{13}{C} pocket.
The s-process nucleosynthesis then proceeds in the pocket with \nuc{13}{C} as the source during the inter-flash phase.
The C22Ne mode then follows during the succeeding helium shell flash.
The nucleosynthesis in the pocket is similar to the C13C mode, except for the low neutron density and the accumulation of neutron poisons during the recurrence.  
The materials are limited by the pocket size and diluted when mixed into the helium flash convection, resulting in smaller efficiencies than the C13C mode.
Moreover, there is an upper limit for the efficiency factor for a given mass of the \nuc{13}{C} pocket. 
The efficiency is much lower than the value predicted by the C13C mode (see \S~\ref{sec:correlations} for details).

\vskip 0.5pc \noindent 
{\bf (II) A Comparison with the Observation:} 

To compare our models with observations, we derived a dilution-free indicator of $\abranc{Z}$, which is given by eq.~(\ref{eq:abrancnZ}).
This indicator is free from the star-to-star variable dilution effect, which is present in $\abra{Z}{H}$ and $\abra{Z}{Fe}$.
This allows us to make a direct comparison between the observed value of $\abranc{Z}$ to the predicted value of $\abranc{Z}$ that reflects the nucleosynthesis in the helium flash convection.

CEMP stars are divided into two subgroups: CEMP-s and CEMP-no stars. In this study, CEMP-no stars are further divided into two subgroups based on $\abranc{Ba}$: Lo-CEMP-no stars with $\abranc{Ba}\ge 0.8$, and Hi-CEMP-no stars with $\abranc{Ba}<0.8$. Hi-CEMP-no stars have larger carbon abundances than Lo-CEMP-no stars.

The largest carbon-normalized enrichments of Sr, Ba, and Pb observed in CEMP stars, including both CEMP-s and CEMP-no stars, are produced by the C13C mode.
The whole range of carbon-normalized enrichments can be reproduced by the combination of the C13C and the subsequent C22Ne modes.
\red{The C22Ne mode produces carbon, which is then dredged-up to the envelope by the TDU.}
On the other hand, the C22Ne mode only produces the lowest end of the carbon-normalized enrichments among CEMP stars.  
The R13C mode can hardly produce sufficient carbon-normalized enrichments, even if the star-to-star variations of \nuc{13}{C} pocket sizes are taken into account. (see \S~\ref{sec:correlations}). 
Most Hi-CEMP-no stars are attributed to the C22Ne mode in massive AGB stars, and also to the C13C mode with larger dilution by the later dredged-up carbon by the C22Ne mode.
Lo-CEMP-no stars are continuously connected to CEMP-s stars with larger Sr/Ba ratios, and to Hi-CEMP-no stars.
It is noted that our models adopt that after the C13C mode, the recurrent TDUs without hydrogen mixing into the He zone could increase surface carbon abundances, which supports such CEMP-no stars with carbon enhancement but without s-process enhancement.    Such a contiguous arrangement is indicative of the common formation
mechanism for all of these subclasses.

\red{
Some CEMP-s stars appear to be single \citep{hansen16a}, which may challenge the wind accretion scenario for these stars. 
However, there is a possibility that such single CEMP-s stars may be nitrogen enhanced metal-poor (NEMP) stars.}
In the wind accretion scenario, NEMP stars form in binaries with intermediate-mass primary stars ($>3\rm{M}_{\odot}$; e.g., Komiya et al. 2007).
In this scenario, NEMP stars should belong to wider binaries than CEMP-s stars, which might make them difficult to be detected as binaries.

Furthermore, the He-FDDM only occurs for low-mass AGB stars ($<3\rm{M}_\odot$; Suda \& Fujimoto 2010), so the s-process materials in NEMP stars should have a different origin from the He-FDDM.
 One possible candidate for the s-process materials in NEMP stars is the very late thermal pulse (VLTP) \citep{herwig11}, which takes place irrespective of the mass of AGB stars.
Moreover, the VLTP might cause the s-process under high neutron densities of $n_n > 10^{12}$ cm${}^{-3}$ (\red{intermediate neutron capture process}).
Therefore, such single NEMP stars might have formed in the VLTP in wide binaries with \red{intermediate-mass} primary AGB stars.

It is also argued that unrealistically high accretion efficiency are required to explain the abundance patterns of CEMP-s and CEMP-r/s \red{stars} \citep{abate13}.
Therefore, it may be even harder to accrete enough material from AGB stars to explain CEMP-no stars in wide binaries in our scenario.
However, even for a wide binary, the accreted mass can be larger when the secondary star is close to the periastron.
This means that it is possible for main sequence secondary stars to accrete enough mass from the primary stars to cover their envelope mass (order of $10^{-3} \rm{M}_\odot$; see e.g., Yamada et al. 2008), although the envelope of the secondary stars will be diluted when it evolves to red giants.

\red{Observations show that there is no differences in the carbon enhancement of CEMP stars between giants and dwarfs, while the dilution effect is observed for lithium abundance \citep{suda11}.
This suggests that the amount of accreted matter in the companions is large enough to enhance carbon in the envelope of CEMP-s and CEMP-r/s giants.}
 
The amount of accreted mass also depends on the wind accretion model.
We adopt carbon dust-driven models, which have lower wind velocities than radiation-driven wind models.
This means that the accreted mass is also larger in dust-driven models than in radiation-driven wind models.
It is necessary to perform detailed hydrodynamical simulations to determine whether this wind model brings enough accreted mass to the secondary star in a wide binary, which is beyond the scope of this paper.

   The present results on the basic properties of these modes provide a foundation for a comprehensive understanding of the abundance distributions of neutron-capture elements in CEMP stars, including the variations among the subclasses, which will be presented in a subsequent paper. 


\begin{ack}
We appreciate helpful discussion by Dr.~W.~Aoki.  
This work was supported in part by JSPS KAKENHI Grant Number, JP23224004, JP24540235, JP25400233, JP15HP7004, JP16K05298, JP16H02168, JP20HP8012, JP22K03688, JP22HP8016, JP23HP8014, and JP19K03931.  
\end{ack}

\begin{table*}
\caption{Model of shell flashes \hspace{10cm}}

\begin{center}
\label{tab:param}
\begin{tabular}{|c|c|c|c|c|c|c|c|}
\hline
Model    &   
$M_{\rm c}{}^{\ast}$ & $r_{\rm c}{}^{\ast\ast}$ & $\log P_{\rm *}{}^{\sharp}$ & $M_{\rm He}$ & $\log T^{\rm max}_{\rm He}{}^{\dagger}$ & $L_{\rm He}^{\rm max}{}^{\ddagger}$  & $N_n^{\rm max}{}^{\S}$ \\ 
 \ name & $(M_{\odot})$ & $ (R_{\odot})$ & (${\rm dyn} / {\rm cm}^2$) & $(\msun)$ & (K) & $(L_\odot)$ &  (cm${}^{-3}$) \\ 
\hline
M840  &  0.60 & 1.51E-2 &  20.37  & 2.3E-2  & 8.404  & 6.2E+6  &  1.65E+12  \\
M844  &  0.70 & 1.41E-2 &  20.34  & 1.4E-2  & 8.442  & 1.5E+7  & 1.09E+12 \\ 
M847  &  0.85 & 1.18E-2 &  20.15  & 3.7E-3  & 8.470   & 6.1E+6 &  1.16E+12 \\ 
M850  &  1.00 & 9.88E-3 &  20.10  & 1.3E-3  & 8.500   & 4.8E+6 & 1.02E+12\\ 
M853  &  1.15 & 7.94E-3 &  20.10  & 4.9E-4  & 8.534   & 4.3E+6 & 8.53E+11 \\ 
\hline
\end{tabular}
\end{center}
${}^{\ast}$ core mass\\
${}^{\ast\ast}$ core radius\\
${}^{\sharp}$ proper pressure\\
${}^{\dagger}$  The maximum temperature achieved in helium flash convection. \\ 
${}^{\ddagger}$ The maximum helium burning rate achieved in helium flash convection.\\  
${}^{\S}$ The maximum neutron density achieved in helium flash convection for the C13C mode with $\dmix=0.03$ and the duration of hydrogen mixing of $10^8$ sec. 
\end{table*}

\appendix

\section{The progress of shell flashes}

The thermal structure of the helium zone during the shell flash was solved analytically by \citet{sugimoto78}. \citet{fujimoto82a,fujimoto99} then formulated a semi-analytical description of the progress of a shell flash based on the theory of shell flashes of finite amplitude, using this analytical solution.  
This semi-analytical description has been applied not only to helium shell flashes, but also to X-ray busters and nova explosions.
It has been shown to be a good representation of their characteristics (Fujimoto et al. 1981; Fujimoto 1982a, Hanawa et al. 1983; Fujimoto et al. 1999; Aikawa et al. 2001, 2004; Nishimura et al. 2009).
   We use the same semi-analytical model to study the \sps\ nucleosynthesis during the helium shell flashes.  

Since the mass in the helium zone is much smaller than the core mass inside the helium zone, i.e., $M_{\rm He} \ll M_c$, then, the pressure, $P_b$, and density, $\rho_b$, at the base of helium burning shell is written in the form
\begin{eqnarray}
P_b =P_{\rm *} f(V_b,N),  \hbox{ and } 
\label{eq:pHe}
\rho_b = V_b ({r_{c} P_*}/{G M_{c} }) f (V_b,N_b),
\label{eq:rhoHe}
\end{eqnarray}
 \red{where $P_*$ is} proper pressure, representing the weight of the overlying helium zone in a flat configuration, i.e., $H_{P, b} \ll r_c$, or $V_b \gg (N_b + 1)$, and hence, $f \simeq 1$, \red{defined by}
\begin{eqnarray}
   P_* = (G M_c / r_c{}^2) (M_{\rm He} / 4 \pi  r_c{}^2),    
\label{eq:proper_pressure}
\end{eqnarray}
 \red{ and  $f$ is }  the flatness parameter, defined by 
\begin{eqnarray}
f (V_b, N) ^{-1}& = & (N+1) \left( \frac{V_b}{N+1} \right)^{N+1} \left[1- (\frac{N+1}{V_b}) \right]^{N-3} \nonumber \\
& \times& \int_0^{(N+1)/V_b} z^N (1-z)^{2-N} dz.   
\label{eq:flatpara}
\end{eqnarray}  
   Here $V_b$ is a homology invariant, representing the ratio of the radius, $r$, to the pressure scale-height, $H_P$, at the base of helium burning shell, i.e., 
\begin{eqnarray}
V_b ={(G M_c / r_c)} / {(P_b / \rho_b)}, 
\label{eq:V}
\end{eqnarray}
   and $N$ is the polytropic index, which may be well approximated by an adiabatic exponent by 
\begin{eqnarray}
N/(N+1) = d \log \rho / d \log P \vert_{\rm ad}.  
\label{eq:N}
\end{eqnarray} 

  As the shell flash develops, $V_b$ decreases and approaches to $N_b +1$, while $N_b + 1$ increases from $\sim 5/2$ to $\sim 4$ \citep{sugimoto78}.  

Consequently the properties of shell flashes are determined by three parameters, the core mass, $M_c$, the core radius, $r_c$, and the proper pressure, $P_*$ (or the mass, $M_{\rm He}$, in the helium zone). 
   The physical quantities can be obtained by solving eqs.~(\ref{eq:pHe}) - (\ref{eq:N}) as a function of entropy, $s_b$, in the burning shell.  
   The time variation of the entropy is then calculated from the energy equation as 
\begin{eqnarray}
\langle T \rangle {ds_b}/{dt}= ({L_{\rm He}}-L_p) / {M_{\rm He}}. 
\label{eq:energy}
\end{eqnarray}
   Here $\langle T \rangle $ is the mass-weighted average of temperature in the helium zone: and $L_{\rm He}$ and $L_{\rm ph}$ are the nuclear energy generation rate and the radiative energy loss rate from the helium zone, respectively.  
   The total energy generation rate in the helium flash convective zone is estimated with the use of thin shell approximation \citep{hayashi62} at 
\begin{eqnarray}
   L_{\rm He}& = & \Delta M_{\rm He} f (V_b, N) \nonumber \\
 & \times & \sum_i  \frac{\varepsilon_{i} (T_b, \rho_b)}{\nu_i \left( \frac{\partial \log T}{\partial \log P} \right)_{\rm ad} + \eta_i \left( \frac{\partial \log \rho}{\partial \log P} \right )_{\rm ad} +1 - \frac{4}{V_b}}, 
\label{eq:lHe}
\end{eqnarray}
   where $\nu_i$ and $\eta_i$ are the temperature and density dependences of the energy generation rate, $\varepsilon_i (\propto T^{\nu_i} \rho^{\eta_i}$).   
   The summation is taken over the reactions with the species of $A \le 34$ with the contribution from the reactions in the extended part neglected.  
   The radiative energy loss rate is given by 
\begin{eqnarray}
L_{\rm ph} = 4 \pi c G (M_c + \Delta M_{\rm He}) (1 - \beta) / \kappa, 
\label{eq:lph}
\end{eqnarray}
   where $1-\beta$ and $\kappa$ are the contribution of radiation pressure and the opacity, respectively.  
   It remains that $ L_{\rm ph} \ll L_{\rm He}$ during the shell flash except for the early stages of igniting shell flash and the end stage of settling into the stable burning.  

\begin{figure}
\includegraphics[width=\columnwidth,bb=0 0 430 300, clip]{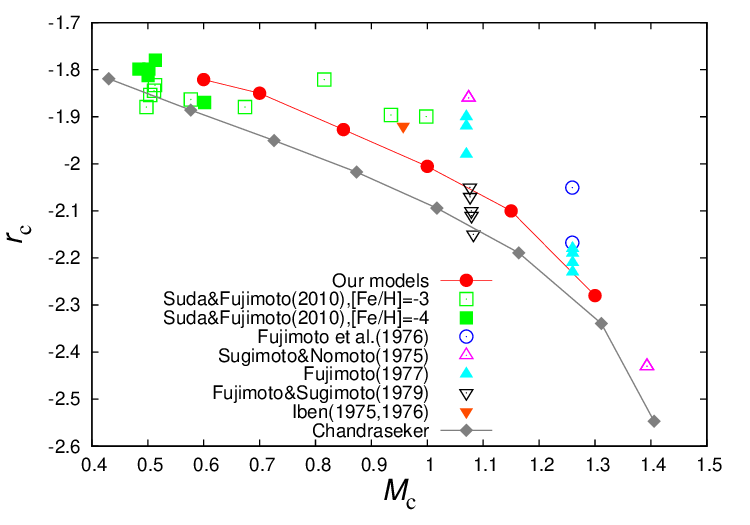}

    \caption{ The mass and the radius of the core interior to the helium burning shell, adopted for our models (filled red circles).   
   Also plotted are the mass and radius of hydrogen-depleted core during the quiescent phase of helium burning taken from the numerical models \citep{suda10,sugimoto75, iben75a,iben76,fujimoto76, fujimoto77,fujimoto79}.  
   Grey diamonds denote the core mass-radius relation of white dwarfs, taken from \citet{chandrasekhar33}. 
    }
 \label{fig:A1massradius}
\end{figure}

\begin{figure}
\includegraphics[width=\columnwidth,bb=75 0 500 323, clip]{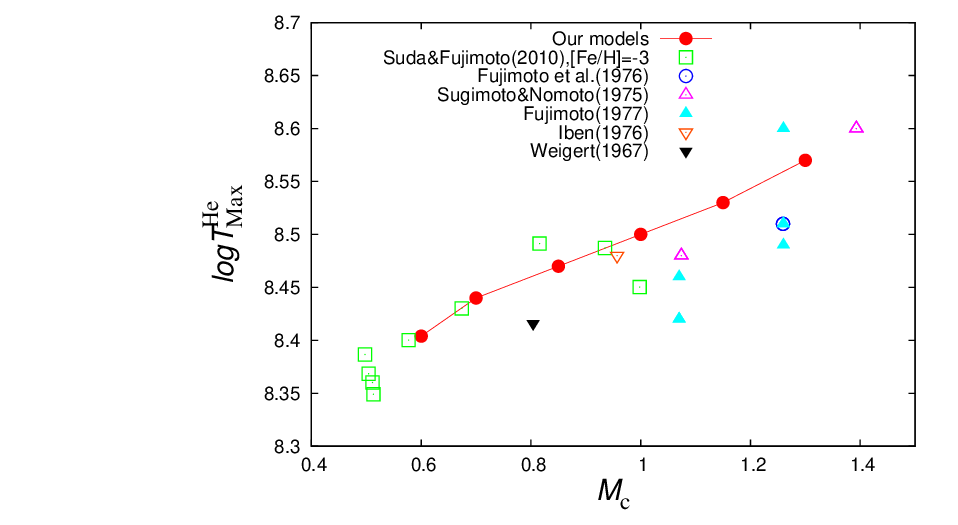}

    \caption{ The maximum temperature at the base of the helium convective zone reached during the shell flashes in our models (filled red circles), are compared with those of the numerical models \citep{suda10,fujimoto77,fujimoto76,iben76,sugimoto75,weigert66}.
   The maximum temperature depends on the proper pressure (or the mass of the helium zone), as well as the mass and radius.
It grows higher during the recurrence in the numerical models as the core cools to delay the ignition \citep{fujimoto79}.
    }
\label{fig:A2maxtemp}
\end{figure}

\begin{figure}
\includegraphics[width=\columnwidth,bb=0 0 415 295, clip]{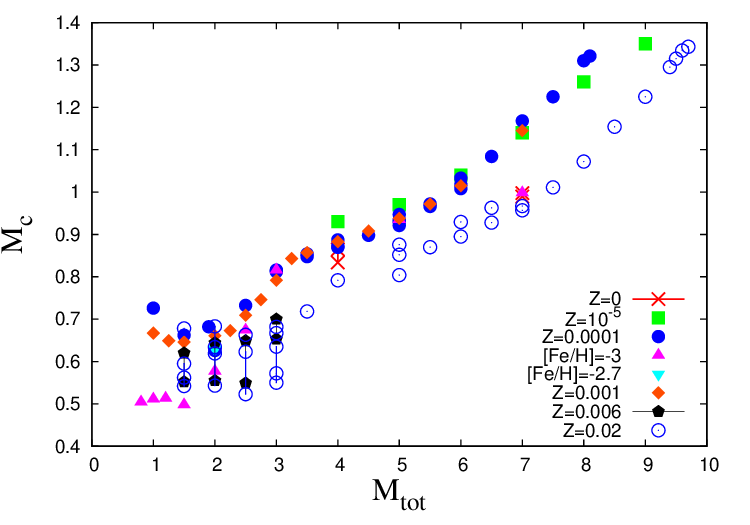}

    \caption{ The core mass of hydrogen-depleted core during the TP-AGB phase against the initial mass of stars, taken from the literature  \citep{weigert66,iben75a,iben76,vassiliadis93,straniero97,chieffi01,iwamoto04,herwig04,karakas07,cristallo09,karakas10,suda10,cristallo11,gilpons13,fishlock14,doherty15}. 
    }
 \label{fig:A3coremass} 
\end{figure}

Table~\ref{tab:param} summarizes the model parameters that we adopted for the shell flashes.
Figure~\ref{fig:flash} shows the time variations in the physical quantities during the shell flashes.   
   Figure~\ref{fig:A1massradius} shows the mass and radius, $M_c$ and $r_c$, of the core interior to the helium burning shell in our models, as well as the comparison with those available from the literature.  
   Figure~\ref{fig:A2maxtemp} shows the comparison of the maximum temperature, $T_{\rm He}^{\rm max}$, reached during the helium shell flashes.  
  Our models well cover the ranges in the numerical computations of shell flashes by \citet{sugimoto78}.  
   The relationship between the core mass and the initial mass can be seen in Figure~\ref{fig:A3coremass}, where the core mass during the TP-AGB phase is plotted against the initial mass of stars from the literature.  



\end{document}